\documentclass[12pt,a4paper,final]{revtex4}

\usepackage{graphicx}% Include figure files
\usepackage{dcolumn}% Align table columns on decimal point
\usepackage{bm}% bold math
\newcommand{\Op}[1]{{\boldsymbol{\mathrm{\hat{#1}}}}}
\newcommand{\Opt}[1]{{\boldsymbol{\mathrm{\tilde{#1}}}}}
\newcommand{\beq}{\begin{equation}}
\newcommand{\eeq}{\end{equation}}
\newcommand{\beqar}{\begin{eqnarray}}
\newcommand{\eeqar}{\end{eqnarray}}
\newcommand{\bea}{\begin{eqnarray}}
\newcommand{\eea}{\end{eqnarray}}
\newcommand{\bcen}{\begin{center}}
\newcommand{\ecen}{\end{center}}

\newcommand{\half}{\frac{1}{2}}
\newcommand{\quarter}{\frac{1}{4}}

\begin{document}

\input epsf.tex    %<-If you need EPS figures to be
                   %  called in {figure} environment for PC
\input epsf.def   %<-If you need EPS figures to be
                   %  called in {figure} environment for Macintosh
%\input psfig.sty
%\jname{..}
%\jyear{2013}
%\jvol{}

%\ARinfo{1056-8700/97/0610-00}

\title{
Quantum heat engines and refrigerators: Continuous devices}

\markboth{Quantum refrigerators}{\textit{ Quantum heat engines and refrigerators}}

\author{Ronnie Kosloff, Amikam Levy}
\affiliation{Institute of Chemistry, the Hebrew University, Jerusalem 91904, Israel}

%\begin{keywords}
%Quantum Thermodynamics, quantum tricycle, quantum amplifier, laser cooling, absolute zero temperature.
%\end{keywords}

\begin{abstract}
Quantum thermodynamics supplies a consistent  description of quantum heat engines and refrigerators
up to the level of a single few level system coupled to the environment. Once the environment is split into three;
a hot, cold and work reservoirs a heat engine can operate. The device converts  the positive gain into power;
where the gain is obtained from population inversion between
the components of the device. Reversing the operation transforms the device into a quantum refrigerator. 
The  quantum tricycle, a device connected by three external leads to three heat reservoirs 
is used as a template for engines and refrigerators. 
The equation of motion for the heat currents and power can be derived from first principle.
Only a global description of the coupling of the device to the reservoirs is consistent 
with the first and second laws of thermodynamics.
Optimisation of the devices leads to a balanced set of parameters where the couplings to the three reservoirs are
of the same order and the external driving field is in resonance.
When analysing refrigerators special attention is devoted to a dynamical version of the third law of thermodynamics.
Bounds on the rate of cooling when $T_c \to 0$ are obtained by optimising the cooling current. 
All refrigerators as $T_c \to 0$ show  universal behavior. The dynamical version
of the third law imposes restrictions on the scaling as $T_c \to 0$ of the relaxation rate $\gamma_c$ 
and heat capacity $c_V$ of the cold bath.

\end{abstract}

\maketitle
\break

 \tableofcontents
 \break
\section{Introduction}
\label{sec:intro}
Our cars, refrigerators, air-conditioners, lasers and power plants are all examples of heat engines. 
The trend toward miniaturisation has not skipped the realm of heat engines
leading to devices on the nano or even on the atomic scale. Typically, in the practical world
all such devices operate far from the  maximum efficiency conditions set by Carnot \cite{carnot}. 
Real heat engines are optimised for powers scarifying efficiency.
This trade-off between efficiency and power is the focus of  "finite time thermodynamics". The field was initiated  
by the seminal paper of Curzon and Ahlboron \cite{curzon75}.
From everyday experience the irreversible phenomena that limits the optimal performance of engines 
can be identified as losses due to friction, heat leaks, and heat transport \cite{salamon01}. 
Is there a  unifying fundamental explanation for these losses? 
Is it possible to trace the origin of these phenomena to quantum mechanics? 
To address these issues the tradition of thermodynamics is followed by the study 
of hypothetical quantum  heat engines and refrigerators. 
Once understood, these models serve as a template for real devices.

{\em Gedanken heat engines} are an integral part of thermodynamical theory.
Carnot in 1824 set the stage by analysing an ideal engine \cite{carnot}. 
Carnot's analysis preceded the systematic  
formulation that led to the first and second laws of thermodynamics \cite{landsberg56}. 
Amazingly, thermodynamics was able to keep its independent status despite 
the development of parallel theories dealing with the same subject matter.
Quantum mechanics overlaps thermodynamics in that it describes the state of matter.
However, in addition, quantum mechanics include a comprehensive description of dynamics.
This suggests that quantum mechanics
can originate a concrete interpretation of the word {\em dynamics} in thermodynamics leading to
a fundamental basis for finite time thermodynamics.

 {\em Quantum thermodynamics} is the study of thermodynamical processes within the context of quantum dynamics. 
Historically, consistency with  thermodynamics led to Planck's law,  the basics of quantum theory. 
Following the ideas of Planck on black body radiation,  Einstein in (1905) quantized the electromagnetic  field \cite{einstein05}.
This paper of Einstein is the birth of quantum mechanics together with quantum thermodynamics.

{\em Quantum thermodynamics} is devoted to unraveling  the intimate connection between the 
laws of thermodynamics and their quantum origin \cite{scovil59,geusic67,k24,k122,partovi89,k156,k169,lloyd,he02,lieb99,bender,kieu04,segal06,bushev06,nori07,casati13,erez08,mahler08,jahnkemahler08,allahmahler08,segal09,he09,nori09,mahlerbook,olshani12,janet13,k281,jarzynski13,allahaverdian2013,erez13}. 
The following questions come to mind:
\begin{itemize}
\item{How do the laws of thermodynamics emerge from quantum mechanics?}
\item{What are the requirements of  a theory to describe quantum mechanics and thermodynamics on a common ground?}
\item{What are the fundamental reasons for  tradeoff between power and efficiency?}
\item{Do quantum devices operating far from equilibrium follow thermodynamical rules?} 
\item{Can quantum phenomena affect the performance of heat engines and refrigerators?}
\end{itemize}
These issues are addressed by analyzing quantum models of heat engines and refrigerators. 
Extreme care has been taken to choose a model which can be analyzed from first principles. 
Two types of models are considered, continuos operating models resembling turbines and discrete four stroke reciprocating engines.
The present review will focus on continuous devices.

\section{The continuous engine}
\label{sec:continuous}

An engine is a device that converts one form of energy to another: heat to work. In this conversion,
part of  heat from the hot bath is ejected to the cold bath limiting the efficiency of power generation. This is the essence
of the second-law of thermodynamics.

A heat engine employs the natural current from a hot to a cold bath to generate power. 
The Carnot engine is a model of such a device. Carnot was able to incorporate the practical knowledge 
on steam engines of his era into a universal 
scientific statement on maximum efficiency \cite{carnot}. 
Out of this insight the laws of thermodynamics were later formulated.
This theme of {\em learning from an example} is typical in thermodynamics and will be employed  to obtain insight
from analysis of  quantum devices.

A continuous engine  operates in an autonomous  fashion attaining steady state
mode of operation. The analysis therefore requires an evaluation of steady sate energy currents.
The operating part of the device is connected simultaneously  to the hot, cold and power output leads. 
The primary macroscopic example is a steam or gas turbine. The primary quantum heat engine is the laser.
These devices share a universal aspect exemplified by the equivalence  of the 3-level laser with the Carnot engine.

Reversing the operation of a heat engine generates a heat pump or a refrigerator. 
We will review the basic principles of quantum continuous 
engines and then invert their operation and study quantum refrigerators.

\subsection{The quantum 3 level system as a heat engine}
\label{subset:3-level}

\begin{figure}
\epsfscale1200		% Figure enlarged to 120%
\center{\includegraphics[height=6cm]{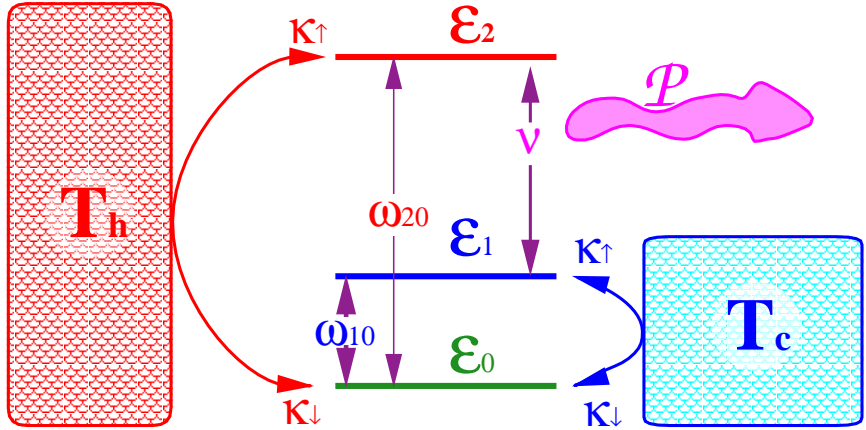}}
\caption{The quantum 3-level amplifier as a Carnot heat engine. 
The system is coupled to a hot bath with temperature $T_h$ and  to a cold bath with temperature $T_c$. 
The output ${\cal P}$ is a radiation field with frequency $\nu$. (In the figure $\hbar=1$).
Power is generated provided there is population inversion between level $\epsilon_2$ and $\epsilon_1$: 
$p_2 > p_1$.
The hot bath equilibrates levels $\epsilon_0$ and $\epsilon_2$ via
the rates $k\uparrow$ and $k\downarrow$ such that $k\uparrow/k\downarrow= \exp(\hbar \omega_{20}/k_B T_h)$.
The efficiency $\eta= \nu/\omega_{20} \le 1-T_c/T_h$. Reversing the direction of operation using power to drive population
from level $\epsilon_1$ to $\epsilon_2$ generates a heat pump then $p_2 < p_1$.}
\label{fig:1}
\end{figure}

A contemporary example of a Carnot engine is the 3-level amplifier.   
The principle of operation is to convert population inversion into output power in the form of  light. 
Heat gradients are employed to achieve this goal. Fig \ref{fig:1} shows its schematic construction.
A hot reservoir characterised by  temperature $T_h$ induces transitions between the ground state
$\epsilon_0$ and the excited state $\epsilon_2$. When equilibrium is reached, the  population ratio between these levels becomes
$$\frac{p_2}{p_0}=e^{-\frac{\hbar \omega_{h}}{k_B T_h}}$$
where $\omega_h \equiv \omega_{20} = (\epsilon_2-\epsilon_0)/\hbar$ is the Bohr frequency and $k_B$ is the Boltzmann constant.
The cold reservoir at temperature $T_c$ couples level $\epsilon_0$ and level $\epsilon_1$ meaning that:
$$\frac{p_1}{p_0}=e^{-\frac{\hbar \omega_{c}}{k_B T_c}}~,$$
where $\omega_c \equiv \omega_{10}=(\epsilon_1-\epsilon_0)/\hbar$.
The amplifier operates by coupling the energy levels $\epsilon_3$ and $\epsilon_2$ to the radiation field  generating an output frequency 
which on resonance is
$\nu = (\epsilon_3-\epsilon_2)/\hbar$. 
The necessary condition for amplification is positive gain or population inversion defined by:
\begin{equation}
G = p_2-p_1 \ge 0~.
\label{eq:gain1}
\end{equation}
From this condition,  by dividing by $p_0$ we obtain 
$ e^{-\frac{\hbar \omega_{h}}{k_B T_h}}-e^{-\frac{\hbar \omega_{c}}{k_B T_c}} \ge 0$,  which leads to:
\begin{equation}
\frac{\omega_c}{\omega_h} \equiv \frac{\omega_{10}}{\omega_{20}} \ge \frac{ T_c}{T_h}~,
\label{eq:ratio}
\end{equation}
The efficiency of the amplifier is defined by the ratio of the output energy $\hbar \nu$ to the energy extracted from the hot reservoir $\hbar \omega_{20}$:
\begin{equation}
\eta_o = \frac{\nu}{\omega_{20}} = 1 - \frac{\omega_c}{\omega_h} ~~.
\label{eq:ottoefic}
\end{equation}
Eq. (\ref{eq:ottoefic}) is termed the quantum Otto efficiency \cite{jahnkemahler08}.
Inserting the positive gain condition Eq. (\ref{eq:gain1}) and Eq. (\ref{eq:ratio}) the efficiency is limited by Carnot  \cite{carnot}:
\begin{equation}
 \eta_o ~\le ~\eta_c ~\equiv ~1 - \frac{T_c}{T_h}~.
 \label{eq:carnot}
 \end{equation}
This result connecting the efficiency of a quantum amplifier to the Carnot efficiency was first obtained by Scuville et al. \cite{scovil59,geusic67}. 

The above description of the 3-level amplifier is based on a static quasi-equilibrium viewpoint.
Real engines which produce power operate far from equilibrium conditions. 
Typically, their performance is restricted by friction, heat transport and heat leaks. 
A dynamical viewpoint is therefore the next required step.

\subsection{The quantum tricycle}
\label{subsec:tricycle}

\begin{figure}
\epsfscale1200		% Figure enlarged to 120%
\center{\includegraphics[height=6cm]{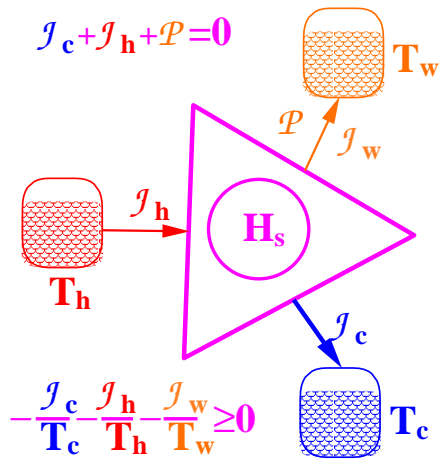}}
\caption{The quantum tricycle: A quantum device coupled simultaneously to a hot, cold and power reservoir. Reversing the heat currents 
constructs a quantum refrigerator.}
\label{fig:2}
\end{figure}

A quantum description enables to incorporate dynamics into thermodynamics. 
The tricycle model is the template for almost all continuous engines Cf. Fig. \ref{fig:2}. This model 
will be employed to incorporate the quantum dynamics of the devices.
Surprisingly very simple models exhibit the same features of engines generating finite power.
Their efficiency at operating conditions is lower than the  Carnot efficiency. 
In addition, heat leaks restrict the preformance meaning that reversible operation is unattainable.

The tricycle engine has a generic structure displayed in Fig. \ref{fig:2}.
\begin{itemize}
\item{The basic model consists of three thermal baths: a hot bath with temperature $T_h$, a cold bath with temperature $T_c$
and a work bath with temperature $T_w$. }
\item{Each bath is connected to the engine via a frequency filter which we will model by three oscillators:
\begin{equation}
\Op H_F = \hbar \omega_h \Op a^{\dagger} \Op a +\hbar \omega_c \Op b^{\dagger} \Op b + \hbar \omega_w \Op c^{\dagger} \Op c~~,
\label{eq:hfilter}
\end{equation}
where $\omega_h$, $\omega_c$ and $\omega_w$ are the filter frequencies on resonance $\omega_w=\omega_h-\omega_c$.}
\item{The device operates as an engine by removing an excitation from the hot bath and generating excitations
on the cold and work reservoirs. In second quantization formalism  the Hamiltonian describing such an interaction becomes:
\begin{equation}
\Op H_I = \hbar \epsilon \left( \Op a \Op b^{\dagger} \Op c^{\dagger} + \Op a^{\dagger} \Op b \Op c \right)~~,
\label{eq:hinteraction}
\end{equation} 
where $\epsilon$ is the coupling strength. 
}
\item{The device operates as a refrigerator by removing an excitation from the cold bath as well as from the work bath
and generating an excitation in the hot bath. The term $\Op a^{\dagger} \Op b \Op c$ in  the Hamiltonian of Eq. (\ref{eq:hinteraction}) describes this action.
(Cf. Section \ref{sec:fridge}).}
 \end{itemize}
Some comments are appropriate:
\begin{enumerate}
\item{
Different types of heat baths can be employed which can include bosonic baths composed of phonons or photons, or
fermonic baths composed of electrons. The frequency filters select from the continuous spectrum of the bath the working
component to be employed in the tricycle. These frequency filters can be constructed also from two-level-systems (TLS) or 
formulated as qubits \cite{k275,popescu10,palao13}. }

\item{The interaction term Eq. (\ref{eq:hinteraction}), is strictly non-linear, incorporating three heat currents simultaneously. 
This crucial fact has important consequences. 
A linear device cannot operate as a heat engine or refrigerator \cite{paz}. 
A linear device is constructed from a network of harmonic oscillators
with linear connections of the type $\hbar \mu_{ij} \left( \Op a_i \Op a^{\dagger}_j  + \Op a^{\dagger}_i \Op a_j \right)~$ with some connections
to harmonic heat baths. In such a device the hottest bath always cools down and the coldest bath always heats up. 
Thus, this construction
can transport heat but not generate power since power is equivalent to transporting heat to an infinitely hot reservoir. 
Another flaw in a linear model is that the different bath modes do not equilibrate with each other. 
A generic bath should equilibrate any system Hamiltonian irrespective of its frequency.}

\item{Many nonlinear interaction Hamiltonians of the type $\Op H_I = \Op A \otimes \Op B \otimes \Op C$ can lead
to a working heat engine. These Hamiltonians can be reduced to the form of Eq. (\ref{eq:hinteraction}) which captures the essence of such interactions.}
\end{enumerate}

The first-law of thermodynamics represents the energy balance of heat currents originating from the three baths and collimating on the system:
\begin{equation}
\frac{dE_s}{dt}= {\cal J}_h + {\cal J}_c +{\cal J}_w ~~.
\label{eq:I-law-t}
\end{equation}
At steady state no heat is accumulated in the tricycle, thus $\frac{dE_s}{dt}= 0$. 
In addition, in steady state the entropy is only generated in the baths, leading to the second-law of thermodynamics:
\begin{equation}
\frac{d}{dt}\Delta {\cal S}_u~=~-\frac{{\cal J}_h}{T_h} - \frac{{\cal J}_c}{T_c} -\frac{{\cal J}_w }{T_w}~\ge~0~~.
\label{eq:II-law-t}
\end{equation}
This version of the second-law is a generalisation of the statement of Clausius; heat does not flow spontaneously from cold to hot bodies \cite{clausius1850}.
When the temperature $T_w \rightarrow \infty$, no entropy is generated in the power bath. 
An energy current with no accompanying entropy production  is equivalent to generating pure power:
${\cal P}={\cal J}_w$, where ${\cal P}$ is the output power.

The evaluation of the currents ${\cal J}_j$  in the tricycle model requires dynamical equations of motion. 
The major assumption is that the total systems is closed
and its dynamics is generated by the global Hamiltonians. 
\begin{equation}
\Op H =\Op H_0 + \Op H_{SB}~.
\label{eq:hamilglobal}
\end{equation}
This Hamiltonian $\Op H_0$ includes the bare system and the heat baths:
\begin{equation}
\Op H_0~~=~~ \Op H_s + \Op H_{H} + \Op H_{C}+ \Op H_{W}~,
\label{eq:hamil-g}
\end{equation}
where the system Hamiltonian  $\Op H_s= \Op H_I +\Op H_h +\Op H_c + \Op H_w$ 
consists of three energy filtering components and an interaction part with an external field. 
The reservoir Hamiltonians are  the hot $\Op H_{H}$, cold $\Op H_{C}$ and work $\Op H_{W}$.
The system-bath interaction Hamiltonian: $\Op H_{SB} = \Op H_{sH}+\Op H_{sC} +\Op H_{sW}$.

A thermodynamical idealisation assumes that the tricycle system and the baths are uncorrelated, meaning that the total state 
of the combined system becomes a tensor product at all times \cite{k281}:
\begin{equation}
\Op \rho = \Op \rho_s \otimes \Op \rho_{H} \otimes \Op \rho_{C} \otimes \Op \rho_{W}~.
\label{eq:rho-t}
\end{equation}
Under these conditions  the  dynamical equations of motion for the tricycle become: 
\begin{equation}
\frac{d}{dt} \Op \rho_s = {\cal L} \Op \rho_s~,
\label{eq:lvn}
\end{equation}
where ${\cal L}$ is the Liouville superoperaor described in terms of the system Hilbert space,
where the reservoirs are  described implicitly.
Within the formalism of quantum open system, $\cal L$ can take the form of the 
Gorini-Kossakowski-Sudarshan-Lindblad (GKS-L) Markovian generator \cite{lindblad76,kossakowski76}.

Thermodynamics is notorious in employing a very small number of variables. 
In equilibrium conditions the knowledge of the Hamiltonian is sufficient.
When deviating from equilibrium additional observables are added. This suggest 
presenting the dynamical generator in Heisenberg form for arbitrary system operators $\Op O_s$:
\begin{equation}
\frac{d}{dt} \Op O_s ={\cal L}^* \Op O_s ~=~ \frac{i }{\hbar}[\Op H_s, \Op O_s] +\sum_k \Op V_k \Op O_s \Op V_k^{\dagger} -\frac{1}{2}\{\Op V_k \Op V_k^{\dagger}, \Op O_s\}~.
\label{eq:lvn-l}
\end{equation}
The operators $\Op V_k$ are system operators still to be determined. 
The task of evaluating the modified system Hamiltonian $\Op H_s$ and the operators $\Op V_k$ is made extremely difficult due to the nonlinear interaction
in Eq. (\ref{eq:hinteraction}). 
Any progress from this point requires a specific description of the heat baths and approximations to deal with the nonlinear terms.

\subsection{The quantum amplifier}
\label{subsec:amplifier}

The quantum amplifier is the most elementary quantum continuous heat engine converting heat to work. 
The purpose is to generate power from a temperature difference between the hot and cold reservoirs.
The output power is described by a time dependent external field.

The general device Hamiltonian is therefore time dependent:
\begin{equation}
\Op H_s(t) = \Op H_0 + \Op V (t)~.
\label{eq:hamil-v}
\end{equation}
The  Markovian master equation in Heisenberg form for the system operator $\Op O_s$
when the system coupled to the hot and cold bath: 
\begin{equation}
\frac{d}{dt}\Op O_s = \frac{i}{\hbar}[ \Op H(t),\Op O_s(t)] + \mathcal{L}^*_h(\Op  O_s(t) )+\mathcal{L}^*_c (\Op O_s(t))~.
\label{ad_markov-1}
\end{equation}
The change in energy of the device is obtained by replacing $\Op O_s$ by $\Op H_s$:
\begin{equation}
\frac{d E_s}{dt} ~=~ \langle {\cal L}_h^* (\Op H_s )\rangle + \langle {\cal L}_c^* (\Op H_s )\rangle +\langle \frac{ \partial \Op H_s}{\partial t} \rangle~.
\label{eq:I-law-t1}
\end{equation}
Equation (\ref{eq:I-law-t1}) can be interpreted as the time derivative of the first law of thermodynamics \cite{sphon,k23,k24,k281} 
based on the Markovian GKS-L generator. 
Power is identified as:
\begin{equation}
{\cal P} = \langle \frac{ \partial H_s}{\partial t} \rangle~,
\label{eq:power}
\end{equation}
and the heat current as:
\begin{equation}
{\cal J}_h = \langle {\cal L}_h^* (\Op H_s )\rangle~~,~~{\cal J}_c = \langle {\cal L}_c^* (\Op H_s )\rangle~.
\end{equation}
The partition between the Hamiltonian and the dissipative part in the  GKS-L generator is not unique \cite{k281}.
A unique derivation of the Master equation based on the weak coupling limit leads to dynamics
which is consistent with the first and second laws of thermodynamics \cite{davis74}.

The template of the tricycle model is employed to describe the dynamics of the amplifier. 
The interaction Hamiltonian is modified to become:
\begin{equation}
\Op H_I (t) = \hbar \epsilon \left(   \Op a ~\Op b^{\dagger} e^{+i \nu t }  + \Op a^{\dagger} \Op b  ~e^{-i \nu t} \right)~~,
\label{eq:h-amplifier}
\end{equation} 
where $\nu \equiv \omega_w$ is the frequency of the time dependent driving field and $\epsilon$ is the coupling amplitude. 
The modification of Eq. (\ref{eq:hinteraction}) eliminates the nonlinearity,
it amounts to replacing the operator $\Op c$ and $\Op c^{\dagger}$ by a c-number. The amplifier output power becomes:
\begin{equation}
\label{eq:power-2}
{\cal P} =  \hbar \epsilon \nu \left(\langle\Op a \Op b^{\dagger} \rangle e^{+i \nu t } - \langle \Op a^{\dagger} \Op b \rangle e^{-i \nu t} \right)~.
\end{equation}

After the nonlinearity has been eliminated the quantum Master equation for the amplifier can be derived from first principles 
based on the weak system bath coupling expansion.
This approximation is a thermodynamic idealisation  equivalent to an isothermal partition between the system and baths \cite{k281}.

The interaction with the baths is given by
\begin{equation}
\Op H_{sb} = \lambda_a (\Op a + \Op a^{\dagger})\otimes \Op R_h + \lambda_b (\Op b+ \Op b^{\dagger})\otimes \Op R_c .
\label{103}
\end{equation}
where $\Op R$ are reservoir operators and $\lambda $ is the small system-bath coupling parameter.
A crucial step has to be performed before this procedure is applied. 
The system Hamiltonian has to be rediagonalized  with the interaction
before the system is coupled to the baths. This diagonalization modifies the frequencies of the system resulting 
in a splitting of the filter frequencies. The prediagonalization is crucial for the master equations 
to be consistent with the second-law of thermodynamics \cite{k114,k122,palao13}.

The main ingredients of the derivation:
\par
I) Transformation to interaction picture. The reservoir coupling operators $\Op R$  
transform according to the free baths Hamiltonian, and the system operators are subject to the unitary 
propagator which under resonance conditions becomes:
\begin{equation}
\Op U_s(t,0) = \mathcal{T}\exp\Bigl\{-\frac{i}{\hbar}\int_0^t \Op H(s) ds\Bigr\}= e^{-\frac{i}{\hbar}\Op H_0 t} e^{-\frac{i }{\hbar}\Op Vt},
\label{101}
\end{equation}
where
\begin{equation}
\Op H_0 = \hbar \omega_h \Op a^{\dagger} \Op a + \hbar \omega_c \Op b^{\dagger} \Op b \ ,\  \Op V =\hbar \epsilon ( \Op a^{\dagger} \Op b +  \Op a \Op b^{\dagger} )\ .
\label{102}
\end{equation}
\par
II) Fourier decomposition of the interaction part. The operators in the interaction picture take the form 
\begin{eqnarray}
\begin{array}{l}
\bf{ \tilde  a}(t)= \Op U_s(t,0)^{\dagger}\Op a \Op U_s(t,0)= \\
e^{\frac{i}{\hbar}\Op Vt} \Bigl[e^{\frac{i}{\hbar}\Op H_0 t}\Op a e^{-\frac{i}{\hbar} \Op H_0 t} \Bigr]e^{-\frac{i}{\hbar}\Op Vt} = 
\cos(\epsilon t) e^{-i\omega_h t}\Op a - i\sin(\epsilon t) e^{-i\omega_h t}\Op b ,
\end{array}
\label{104}
\end{eqnarray}
and a similar equation for ${\bf{ \tilde  b}}(t)$.
The Fourier decomposition becomes:
\begin{equation}
\bf{\tilde  a} (t)=\frac{1}{\sqrt{2}} (e^{-i(\omega_{h}^{+}) t} \bf{\tilde d}_+ + e^{-i(\omega_{h}^{-} )t}\bf{\tilde d}_-)
\end{equation}
and
\begin{equation}
\bf{\tilde b}(t)=\frac{1}{\sqrt{2}} (e^{-i(\omega_c^+)t}\bf{\tilde d}_+ - e^{-i(\omega_c^- )t}\bf{\tilde d}_-  )
\end{equation}
where $\bf{\tilde d}_+=\frac{\bf{\tilde a}+ \bf{\tilde b}}{\sqrt{2}}$, $\bf{\tilde d}_-=\frac{\bf{\tilde a}-\bf{\tilde b}}{\sqrt{2}}$ and 
$\omega_{h(c)}^{\pm}=(\omega_{h(c)} \pm \epsilon)$. Similarly $\bf{ \tilde a}^{\dagger}(t), \bf{\tilde b}^{\dagger}(t)$ are evaluated.

III) Performing the system-bath weak coupling approximation. 
This approximation involves averaging over fast oscillating terms with typical frequencies $ \sim 2\omega_c , 2\omega_h , 2\epsilon$. 
This approximation is restricted to conditions  that the relaxation time of the open system 
is much longer than the intrinsic time scale $\omega^{-1}_c,\omega^{-1}_h$ and $\epsilon^{-1}$. 
Thus, terms oscillating rapidly over the relaxation time average out. Such equations were derived in Ref.\cite{k275}.
If the coupling to the external field is weak, such that $\omega_h,\omega_c\gg \epsilon$ and the relaxation time of the 
open system is comparable with $\epsilon^{-1}$, the derivation is modified accordingly. 
In this case there is no justification to neglect terms oscillating with frequency $\sim \epsilon$. 
Keeping such terms, the total time-dependent (interaction picture) generator has the form:
\begin{equation}
\mathcal{L}(t)= \mathcal{L}_h^{(+)} + \mathcal{L}_h^{(-)}+\mathcal{L}_c^{(+)}+\mathcal{L}_c^{(-)} ~,
\label{106}
\end{equation}
where 
\begin{eqnarray}
\begin{array}{ll}
\label{eq:generators}
{\cal L}_{h}^{(+)} \Op \rho= \frac{1}{4}\gamma_{h}^{(+)} \Bigl( [{\bf{\tilde d}_+} ,\Op \rho {\bf{\tilde d}_+^{\dagger}}]+
[{\bf{\tilde d}_- },\Op \rho {\bf{\tilde d}_+^{\dagger}}]e^{i2\epsilon t} 
+ e^{-\beta_{h}\omega_{h}^{+}}\left([{\bf{\tilde d}_+^{\dagger} },\Op \rho {\bf{ \tilde d}_+}] + 
[{\bf{\tilde d}_-^{\dagger}} ,\Op \rho {\bf{\tilde d}_+}]e^{-i2\epsilon t}\right) +h.c\Bigr) \\
\\
{\cal L}_{c}^{(+)}\Op  \rho= \frac{1}{4}\gamma_{c}^{(+)} \Bigl( [{\bf{\tilde d}_+ },\Op \rho {\bf{ \tilde d}_+^{\dagger}}]-
[{\bf{\tilde d}_-} ,\Op \rho {\bf{\tilde d}_+^{\dagger}}]e^{i2\epsilon t} + e^{-\beta_{c}\omega_{c}^{+}}\left([{\bf{\tilde d}_+^{\dagger}} ,\Op \rho {\bf{\tilde d}_+}] - 
[{\bf{\tilde d}_-^{\dagger}} ,\Op \rho {\bf{\tilde d}_+}]e^{-i2\epsilon t}\right) +h.c\Bigr) \\
\\
{\cal L}_{h}^{(-)} \Op \rho= \frac{1}{4}\gamma_{h}^{(-)} \Bigl( [{\bf{\tilde d}_- },\Op \rho {\bf{\tilde d}_-^{\dagger}}]+[{\bf{\tilde d}_+ },\Op \rho {\bf{\tilde d}_-^{\dagger}}]e^{-i2\epsilon t} + e^{-\beta_{h}\omega_{h}^{-}}\left([{\bf{\tilde d}_-^{\dagger}} ,\Op \rho {\bf{\tilde d}_-}]+
[{\bf{\tilde d}_+^{\dagger}} ,\Op \rho {\bf{\tilde d}_-}]e^{i2\epsilon t}\right) + h.c\Bigr) \\
\\
{\cal L}_{c}^{(-)} \Op \rho= \frac{1}{4}\gamma_{c}^{(-)} \Bigl( [{\bf{\tilde d}_-} ,\Op \rho {\bf{\tilde d}_-^{\dagger}}]
-[{\bf{\tilde d}_+} ,\Op \rho {\bf{\tilde d}_-^{\dagger}}]e^{-i2\epsilon t} + e^{-\beta_{c}\omega_{c}^{-}}\left([{\bf{\tilde d}_-^{\dagger}} ,\Op \rho {\bf{\tilde d}_-}]-
[{\bf{\tilde d}_+^{\dagger} },\Op \rho {\bf{ \tilde d}_-}]e^{i2\epsilon t}\right) + h.c\Bigr) \\
\end{array}
\end{eqnarray} 
and the inverse temperature is $\beta = \hbar/k_B T$.
The relaxation rates $\gamma_{h(c)}^{(\pm)}=\gamma_{h(c)}(\omega_{h(c)}\pm \epsilon)$, have the structure:
\begin{equation}
 \gamma(\omega)=\lambda^2\int^{\infty}_{-\infty}e^{i\omega t} Tr(\Op \rho_R e^{\frac{i}{\hbar}\Op H_R t}\Op R e^{-\frac{i}{\hbar}\Op H_R t}\Op R)dt 
\end{equation}
and can be calculated explicitly for different types of heat baths.
\par
Note that if we neglect the time dependent terms in Eq. (\ref{eq:generators}) the master equation derived in Ref. \cite{k275} is recovered.
The generators in Eq.(\ref{eq:generators}) are not in GKS-L completely positive semigroup form. To correct this flaw,
off-diagonal terms are added to the master equation recovering the first standard form, i.e. 
${\cal L}_D \Op \rho =\sum_{i,j} C_{i,j} \left( \Op F_i \Op \rho \Op F_j^{\dagger}-\half \{ \Op F_j^{\dagger} \Op F_i,\Op \rho \} \right)$.
Next, we insure that these terms will not contribute twice by rescaling the kinetic coefficients.  
In order to verify that the map is completely positive, it is sufficient  that the matrix $(C_{i,j})$  
is positive definite. The modified GKS-L master equations now become:
\begin{eqnarray}
\begin{array}{ll}
\label{eq:generators2}
{\cal L}_{h}^{(\pm)} \Op \rho= \frac{1}{2}\gamma_{h}^{(\pm)} \Bigl(  {\bf{\tilde d}_{\pm}}\Op \rho {\bf{\tilde d}_{\pm}^{\dagger}}
 -\half \{ {\bf{\tilde d}^{\dagger}_{\pm}}{\bf{\tilde d}_{\pm}},\Op \rho \}  
 +e^{-\beta_h \omega_h^{\pm}} \left( {\bf{\tilde d}_{\pm}^{\dagger}} \Op \rho {\bf{\tilde d}_{\pm}}
  -\half \{ {\bf{\tilde d}_{\pm}}{\bf{\tilde d}_{\pm}^{\dagger}},\Op \rho \} \right)  \Bigr) \\
\\
~~~~~~~ + \quarter \gamma_{h}^{(\pm)} \Bigl(  {\bf{\tilde d}_{-}} \Op \rho {\bf{\tilde d}_{+}^{\dagger}} 
-\half \{ {\bf{\tilde d}^{\dagger}_{+}}{\bf{\tilde d}_{-}},\Op \rho \}  
+e^{-\beta_h \omega_h^{\pm}} \left( {\bf{\tilde d}_{+}^{\dagger}} \Op \rho {\bf{\tilde d}_{-}}
 -\half \{{\bf{\tilde  d}_{-}}{\bf{\tilde d}_{+}^{\dagger}},\Op \rho \} \right)   \Bigr)e^{i2\epsilon t}\\
\\
~~~~~~~ + \quarter \gamma_{h}^{(\pm)} \Bigl(  {\bf{\tilde d}_{+}}\Op \rho {\bf{\tilde d}_{-}^{\dagger}} 
-\half \{ {\bf{\tilde d}^{\dagger}_{-}}{\bf{\tilde d}_{+}},\Op \rho \}  
+e^{-\beta_h \omega_h^{\pm}} \left( {\bf{\tilde d}_{-}^{\dagger}}\Op \rho {\bf{\tilde d}_{+}} 
-\half \{ {\bf{\tilde d}_{+}}{\bf{\tilde d}_{-}^{\dagger}},\Op \rho \} \right)   \Bigr)e^{-i2\epsilon t}\\
\\
{\cal L}_{c}^{(\pm)} \Op \rho= \frac{1}{2}\gamma_{c}^{(\pm)} \Bigl(  {\bf{\tilde d}_{\pm}}\Op \rho {\bf{\tilde d}_{\pm}^{\dagger}}
 -\half \{ {\bf{\tilde d}^{\dagger}_{\pm}}{\bf{\tilde d}_{\pm}},\Op \rho \}  
 +e^{-\beta_c \omega_c^{\pm}} \left( {\bf{\tilde d}_{\pm}^{\dagger}}\Op \rho {\bf{\tilde d}_{\pm}} 
 -\half \{ {\bf{\tilde d}_{\pm}}{\bf {\tilde d}_{\pm}^{\dagger}},\Op \rho \} \right)  \Bigr) \\
\\
~~~~~~~ - \quarter \gamma_{c}^{(\pm)} \Bigl(  {\bf{\tilde d}_{-}}\Op \rho {\bf{\tilde d}_{+}^{\dagger}} 
-\half \{ {\bf{\tilde d}^{\dagger}_{+}}{\bf{\tilde d}_{-}},\Op \rho \}  
+e^{-\beta_c \omega_c^{\pm}} \left( {\bf{\tilde d}_{+}^{\dagger}}\Op \rho {\bf{\tilde d}_{-}} 
-\half \{ {\bf{\tilde d}_{-}}{\bf{\tilde d}_{+}^{\dagger}},\Op \rho \} \right)   \Bigr)e^{i2\epsilon t}\\
\\
~~~~~~~ - \quarter \gamma_{c}^{(\pm)} \Bigl(  {\bf{\tilde d}_{+}}\Op \rho {\bf{\tilde d}_{-}^{\dagger}}
 -\half \{ {\bf{\tilde d}^{\dagger}_{-}}{\bf{\tilde d}_{+}},\Op \rho \}  
 +e^{-\beta_c \omega_c^{\pm}} \left( {\bf{\tilde d}_{-}^{\dagger}} \Op \rho {\bf{\tilde d}_{+}} 
 -\half \{ {\bf{\tilde d}_{+}}{\bf{\tilde d}_{-}^{\dagger}},\Op \rho \} \right)   \Bigr)e^{-i2\epsilon t}
\end{array}
\end{eqnarray} 
Note that the rotating term $e^{-i2\epsilon t}$ can be absorbed in ${\bf {\tilde d}}$ by a second rotation of the frame. 
The derivation of the Master equation, Eq. (\ref{eq:generators2}), for a driven system is a delicate task. 
The pre-diagonalization step ensures  consistency with the second-law \cite{k114,k122,alicki2013}.

\subsubsection{Solving the equations of motion for the engine}

In thermodynamic tradition, the engine is well described by a small set of observables. 
They in turn are represented by operators defining the heat currents in the engine.
To exploit this property the Hamiltonian is transformed to the interaction frame: 
\begin{equation}
 \Op H_I(t)=\Op U_s^{\dagger}(t,0)\Op H(t)\Op U_s(t,0)=\hbar \frac{\omega_h+\omega_c}{2} \Op W + \hbar \frac{\omega_h-\omega_c}{2} \Op X + \hbar \epsilon \Op Z 
\end{equation}
where the operators are closed to commutation relations, forming the SU(2) Lie algebra:\\
$\Op W=({\bf{\tilde d}^{\dagger}_+}{\bf{\tilde d}_+} +{\bf{\tilde d}^{\dagger}_-}{\bf{\tilde d}_-})
~~~,~~~\Op X=({\bf{\tilde d}^{\dagger}_+}{\bf{\tilde d_-}}e^{i2\epsilon t} 
+{\bf{\tilde d}^{\dagger}_-}{\bf{\tilde d}_+} e^{-i2\epsilon t})$,\\
$\Op Y=i({\bf{\tilde d}^{\dagger}_+}{\bf{\tilde d}_-}e^{i2\epsilon t} -{\bf{\tilde d}^{\dagger}_-}{\bf{\tilde d}_+} e^{-i2\epsilon t})~~~, 
~~~\Op Z=({\bf {\tilde d}^{\dagger}_+}{\bf{\tilde d}_+} -{\bf{\tilde d}^{\dagger}_-}{\bf{\tilde d}_-}) $.\\
The dynamical description of the engine is completely determined by the expectation values of the operators constituting the SU(2) algebra.
The equation of motion for these operators which are closed to this set follow Eq. (\ref{eq:generators2}) in Heinsenberg form:
\begin{eqnarray}
\begin{array}{ll}
\frac{d \Op W}{dt}= -\quarter \Gamma_T \Op W -\quarter(\Gamma^+ -\Gamma^-)\Op Z -\quarter (\Gamma_h -\Gamma_c)\Op X + \half (\Gamma_h^+N_h^+ +\Gamma_h^-N_h^- +\Gamma_c^+N_c^+ +\Gamma_c^-N_c^-)\\
\\
\frac{d\Op X}{dt}= -\quarter \Gamma_T \Op X -\quarter (\Gamma_h -\Gamma_c)\Op W + \half (\Gamma_h^+N_h^+ +\Gamma_h^-N_h^- -\Gamma_c^+N_c^+ -\Gamma_c^-N_c^-)+2\epsilon \Op Y\\
\\
\frac{d\Op Y}{dt}=-\quarter\Gamma_T \Op Y -2\epsilon \Op X\\
\\
\frac{d\Op Z}{dt}= -\quarter \Gamma_T \Op Z -\quarter(\Gamma^+ -\Gamma^-)\Op W + \half (\Gamma_h^+N_h^+ -\Gamma_h^-N_h^- +\Gamma_c^+N_c^+ -\Gamma_c^-N_c^-)~,\\
\end{array}
\end{eqnarray} 
where the equilibrium populations of the dressed filter operators and  heat transport coefficients become:
\begin{eqnarray}
\begin{array}{ll}
N_{h(c)}^{\pm}= \frac{1}{\exp\left({\frac{\hbar \omega_{h(c)}^{\pm}}{k_B T_{h(c)}}}\right)-1}~~,~~
\Gamma_{h(c)}^{\pm}=\gamma_{h(c)}^{\pm} (1-\exp \left({-\frac{\hbar \omega_{h(c)}^{\pm}}{k_B T_{h(c)}}}\right)~)
\end{array}
\end{eqnarray} 
with $\omega_{h(c)}^{\pm}=\omega_{h(c)}\pm \epsilon$
and
\begin{eqnarray}
\begin{array}{ll}
\Gamma_T = \Gamma_h^+ +\Gamma_h^- +\Gamma_c^+ +\Gamma_c^- 
~,~
\Gamma^{\pm} =\Gamma_h^{\pm} +\Gamma_c^{\pm}
~,~
\Gamma_{h(c)}=\Gamma_{h(c)}^+ +\Gamma_{h(c)}^-
\end{array}~.
\end{eqnarray} 
A simplifying limit is obtained when the relaxation rate of the upper and lower manifold
are approximately equal: $ \Gamma_h^+ =\Gamma_h^- \equiv \kappa_h$ and $ \Gamma_c^+ =\Gamma_c^- \equiv \kappa_c$. 
For a typical harmonic bath where $ \Gamma(\omega) \sim \omega^d$, and $d$ 
stands for the bath dimension, this is a good approximation  ($\omega_h,\omega_c\gg \epsilon$).

The equations of motion for the amplifier relax to a steady state operational mode. 
The expectation values for the operators in steady state $ \frac{ d\Op X}{dt}= \frac{ d\Op Y}{dt}= \frac{ d\Op Z}{dt}= \frac{ d\Op W}{dt}=0 ~$ become:
\begin{eqnarray}
\label{eq:yyy}
\begin{array}{ll}
\left< \Op X \right>= \frac{\kappa_h \kappa_c(N_h^+ +N_h^- -N_c^+ -N_c^-)}{2(4\epsilon^2 + \kappa_h \kappa_c)}\\
\\
\left< \Op Y \right>= - \frac{2\epsilon (N_h^+ +N_h^- -N_c^+ -N_c^-)}{4\epsilon^2 + \kappa_h \kappa_c}\left( \frac{\kappa_h \kappa_c}{\kappa_h+ \kappa_c}\right)\\
\\
\left< \Op Z \right>= \frac{(N_h^+ -N_h^-)\kappa_h + (N_c^+ -N_c^-)\kappa_c}{\kappa_h +\kappa_c}\\
\\
\left< \Op W \right>=\frac{(N_h^+ +N_h^- +N_c^+ +N_c^-)\kappa_h \kappa_c}{2(4\epsilon^2 + \kappa_h \kappa_c)} + \frac{4\lambda^2 ((N_h^+ +N_h^-)\kappa_h + (N_c^+ +N_c^-)\kappa_c) }{(4\epsilon^2 + \kappa_h \kappa_c)(\kappa_h +\kappa_c)} ~.
\end{array}
\end{eqnarray}
The commutator $[\Op Y, \Op H_I] \ne 0$ therefore $\Op Y$ is related to the non diagonal elements
of the Hamiltonian which define the  coherence between energy levels. 

The solution of the equation of motion lead to the thermodynamical observables at steady state conditions, 
the power and heat flows become:
\begin{eqnarray}
\begin{array}{ll}
\label{eq:flows}
\mathcal{P}=\hbar \epsilon \nu \left< \Op Y \right> = -\frac{2\hbar (\omega_h-\omega_c)  \epsilon^2  G_1}{4\epsilon^2 + \kappa_h \kappa_c}\left( \frac{\kappa_h \kappa_c}{\kappa_h+ \kappa_c}\right)\\
\\
\mathcal{J}_h= \mathcal{L}_h^{\dagger}(\Op H_I)= \left(\frac{\hbar \epsilon  G_2}{2} + \frac{2\hbar \omega_h \epsilon^2 G_1}{4\epsilon^2 +\kappa_h \kappa_c}\right)\left( \frac{\kappa_h \kappa_c}{\kappa_h+ \kappa_c}\right) \\
\\
\mathcal{J}_c= \mathcal{L}_c^{\dagger}(\Op H_I)= -\left(\frac{\hbar \epsilon  G_2}{2} + \frac{2\hbar \omega_c \epsilon^2 G_1}{4\epsilon^2 +\kappa_h \kappa_c}\right)\left( \frac{\kappa_h \kappa_c}{\kappa_h+ \kappa_c}\right)\\
\end{array}
\end{eqnarray}
where the generalised gain becomes:
\begin{eqnarray}
\begin{array}{ll}
G_1=(N_h^+ +N_h^-) -(N_c^+ + N_c^-)\\
G_2=(N_h^+ -N_h^-) - (N_c^+ -N_c^-)
\end{array}
\label{eq:gain}
\end{eqnarray}
The first law of thermodynamics is satisfied
such that ${\cal J}_h + {\cal J}_c + {\cal P}=0$ as well as the second law:
$-\frac{{\cal J}_h}{T_h}-\frac{{\cal J}_c}{T_c} \ge 0$. 
The power $\cal P$ is proportional to the expectation value of the coherence $\left< \Op Y \right>$.
As a consequence additional pure dephasing originating from external noise will degrade the power. Such noise generated by a Gaussian 
random process is described by the generator ${\cal L}_D= -\gamma^2[\Op H_I,[\Op H_I, \bullet ]]$  \cite{gorini76}.

In the regime where the external driving amplitude is larger than the heat conductivity $\epsilon^2 \gg \kappa_h \kappa_c$, 
Eq.(\ref{eq:flows}) converges to the results of Ref. \cite{k275}.
\begin{eqnarray}
\begin{array}{ll}
\label{eq:flows2}
\mathcal{P}=\hbar \epsilon\nu \left< \Op Y \right> = -\half \hbar \nu  G_1\left( \frac{\kappa_h \kappa_c}{\kappa_h+ \kappa_c}\right)\\
\\
\mathcal{J}_h= \mathcal{L}_h^{\dagger}(\Op H_I)= \half \hbar \left(\epsilon G_2 +  \omega_h  G_1\right) \left( \frac{\kappa_h \kappa_c}{\kappa_h+ \kappa_c}\right) \\
\\
\mathcal{J}_c= \mathcal{L}_c^{\dagger}(\Op H_I)= -\half \hbar  \left(\epsilon  G_2 +   \omega_c  G_1\right)\left( \frac{\kappa_h \kappa_c}{\kappa_h+ \kappa_c}\right)\\
\end{array}
\end{eqnarray}

\begin{figure}
\epsfscale1200		% Figure enlarged to 120%
\center{\includegraphics[height=8cm]{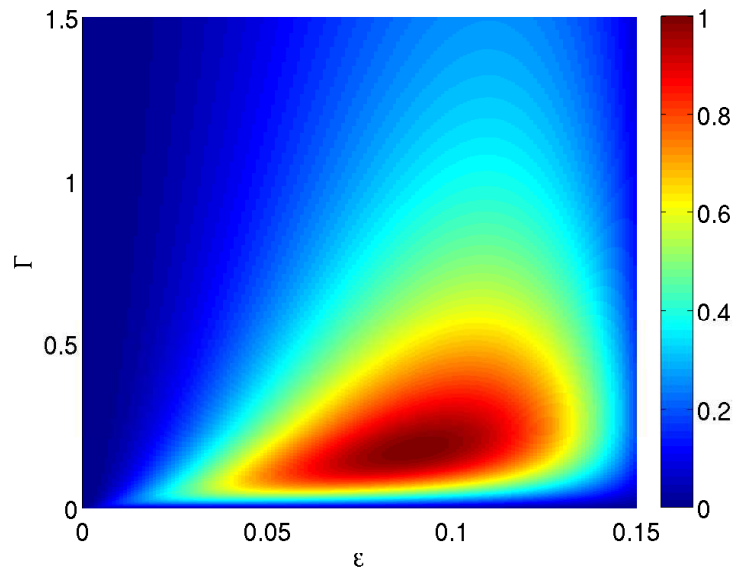}}
\caption{The normalized power  ${\cal P}/{\cal P}_{max}$
as a function of the coupling amplitude to the external field $\epsilon$ and the heat conductivity $\Gamma$.
A clear maximum is obtained for $\Gamma= 2 \epsilon$.}
\label{fig:3}
\end{figure}

Returning to Eq. (\ref{eq:flows}),
optimal power is obtained when the heat conductances from the hot and cold bath are balanced: 
$\Gamma \equiv \kappa_c =\kappa_h$,
then the power and the heat flows from the reservoirs become:
\begin{eqnarray}
\begin{array}{ll}
\mathcal{P}=\hbar \epsilon \nu \left<\Op  Y \right> = -\frac{\hbar \nu  \epsilon^2 \Gamma G_1}{4\epsilon^2 + \Gamma^2}\\
\\
\mathcal{J}_h= \mathcal{L}_h^{\dagger}(\Op H_I)= \frac{\hbar \epsilon \Gamma G_2}{4} + \frac{\hbar \omega_h \epsilon^2 \Gamma G_1}{4\epsilon^2 +\Gamma^2}\\
\\
\mathcal{J}_c= \mathcal{L}_c^{\dagger}(\Op H_I)= -\frac{\hbar \epsilon \Gamma G_2}{4} - \frac{\hbar \omega_c \epsilon^2 \Gamma G_1}{4\epsilon^2 +\Gamma^2}\\
\end{array}
\label{eq:pjj}
\end{eqnarray}
Figure \ref{fig:3} shows the power as a function of the heat conductivity coefficient $\Gamma$
and the coupling to the external field $\epsilon$. A clear global maximum is obtained for $\Gamma=2 \epsilon$.

The efficiency of the amplifier,  is defined as $\eta= - \frac{\cal P}{{\cal J}_h}$.
For the present model it  becomes:
\begin{equation}
\eta = \frac{\nu}{\omega_h+\frac{(\Gamma^2+\epsilon^2)G_2}{4 |\epsilon | G_1}} ~.
\label{eq:eta}
\end{equation}
In the limit of $\epsilon \rightarrow 0$ and $\Gamma \rightarrow 0$ the efficiency approaches the Otto value:
$\eta_o = 1-\frac{\omega_c	}{\omega_h}$ and for zero gain, $\frac{\omega_c}{\omega_h}=\frac{T_c}{T_h}$,
we obtain the Carnot limit  $\eta_c = 1- \frac{T_c}{T_h}$.

\begin{figure}
\epsfscale1200		% Figure enlarged to 120%
\center{\includegraphics[height=6cm]{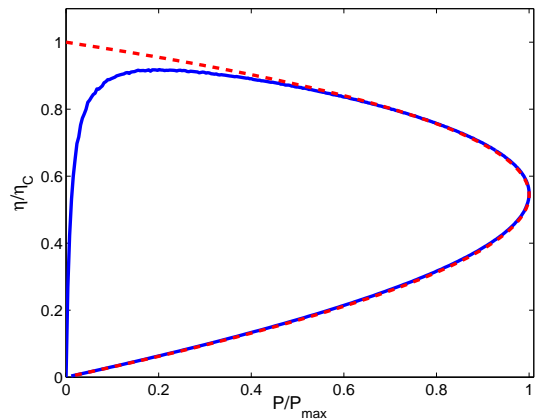}}
\caption{The normalised efficiency $\eta/\eta_c$ vs. the normalised power  ${\cal P}/ {\cal P}_{max}$
for fixed $\epsilon$ and $\Gamma$ and $\omega_c$ while varying $\omega_h$. The blue line is for finite $\Gamma$
while the dashed red line is  optimised at each point $\Gamma=2 \epsilon$. In this case $\epsilon \ll 1$.}
\label{fig:4}
\end{figure}

For finite fixed $\Gamma$ the efficiency in the limit $\epsilon \rightarrow 0$ becomes:
\begin{equation}
 \eta=\frac{\nu}{\omega_h + \frac{\Gamma^2\left(T_h(1-\cosh(\frac{\omega_h}{T_h}))-T_c(1-\cosh(\frac{\omega_c}{T_c}))\right) }{4 T_h T_c\left( \sinh(\frac{\omega_h}{T_h}) -\sinh(\frac{\omega_c}{T_c}) +\sinh(\frac{\omega_c}{T_c}-\frac{\omega_h}{T_h})\right) }}~.
 \label{eq:eficency2}
 \end{equation}
 Examining Eq. (\ref{eq:eficency2}) shows that the Carnot limit is unattainable. 
 When the Otto efficiency  approached the Carnot limit $\frac{\omega_c}{\omega_h}=\frac{T_c}{T_h}$
 the amplifiers efficiency becomes zero. 
 This comparison between the two limits can be seen in Fig. \ref{fig:4}.
 
\begin{figure}
\epsfscale1200		% Figure enlarged to 120%
\center{\includegraphics[height=8cm]{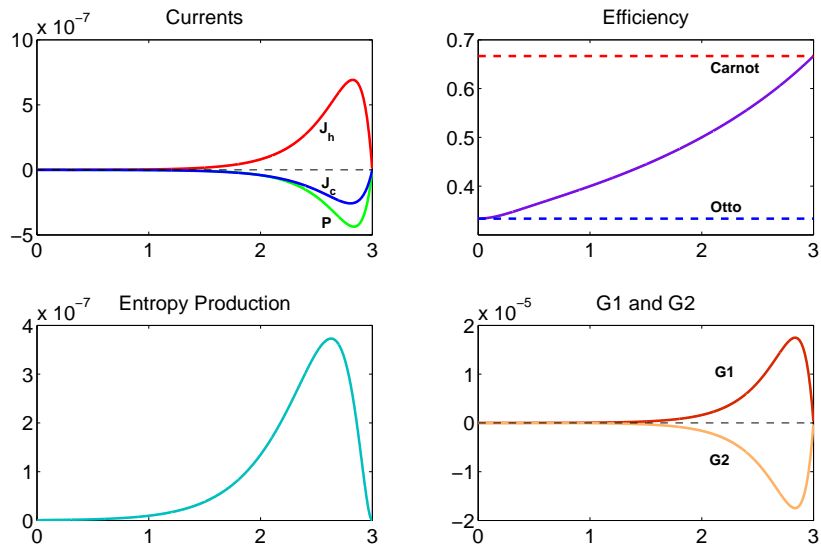}}
\caption{Top left: The heat ${\cal J}_c, {\cal J}_h$ and power ${\cal P}$ currents as a function of $\epsilon$. 
Top Right: Efficiency $\eta$ as a function of $\epsilon$.
Bottom left: Entropy production as a function of $\epsilon$. Bottom right: The gain $G_1$ and $G_2$ as a function of $\epsilon$.
The harmonic tricycle engine operates in the limit of low temperature. $k_B T_c=0.1$ and $k_B T_h =0.3$, $\hbar \omega_c=4.$ and $\hbar \omega_h=6$ and
$\Gamma =0.05$.}
\label{fig:4b}
\end{figure}

\begin{figure}
\epsfscale1200		% Figure enlarged to 120%
\center{\includegraphics[height=8cm]{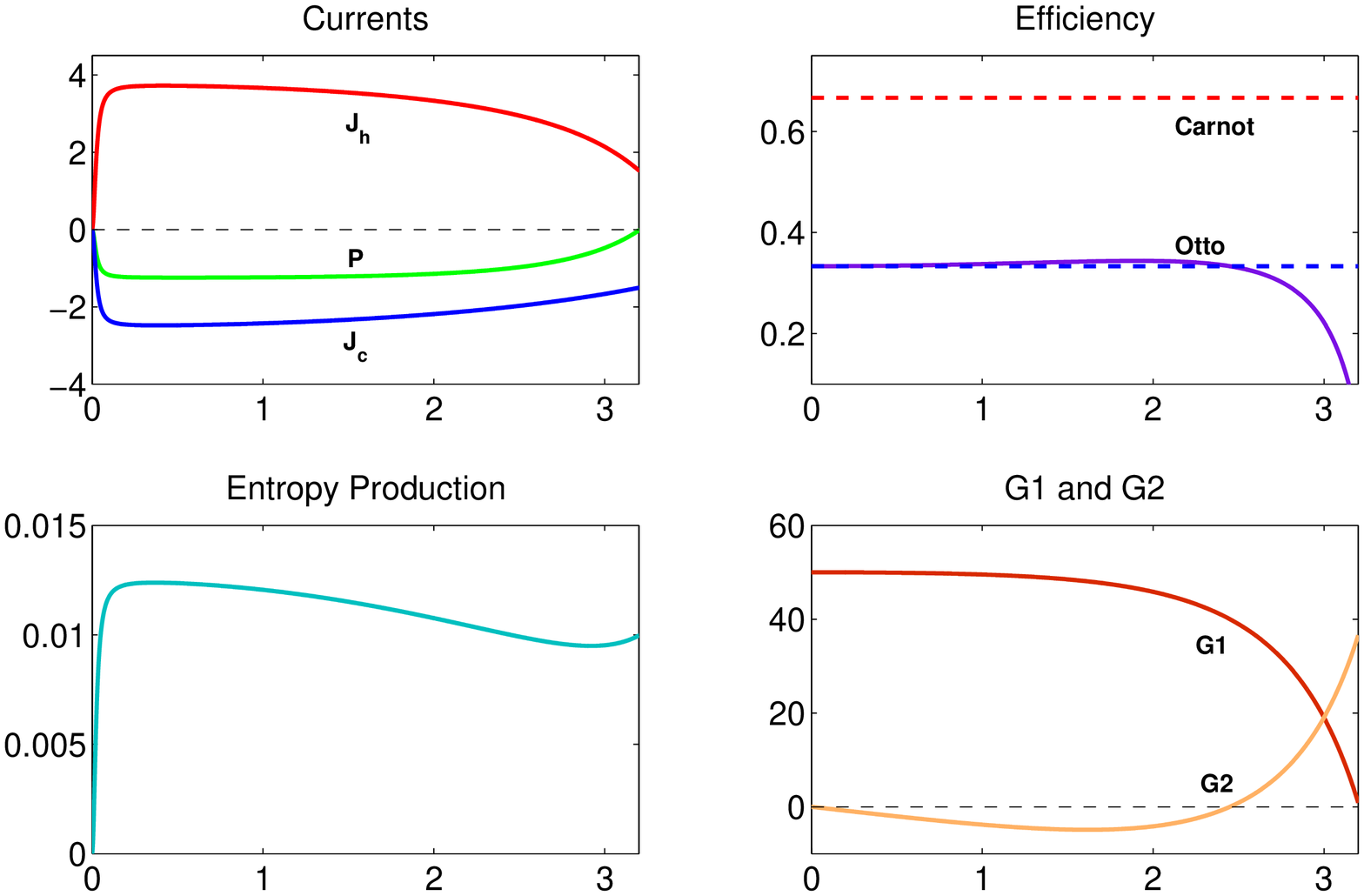}}
\caption{Top left: The heat ${\cal J}_c, {\cal J}_h$ and power ${\cal P}$ currents as a function of $\epsilon$. Top Right: Efficiency $\eta$ as a function of $\epsilon$.
Bottom left: Entropy production as a function of $\epsilon$. Bottom right: The gain $G_1$ and $G_2$ as a function of $\epsilon$.
The harmonic tricycle engine operates in the limit of high  temperature. $k_B T_c=100$ and $k_B T_h =300$, $\hbar \omega_c=4.$ and $\hbar \omega_h=6$ and
$\Gamma =0.05$.}
\label{fig:4c}
\end{figure}

The efficiency of Eq. (\ref{eq:eta}) can be either smaller or greater than the Otto efficiency depending on the sign of $G_2$.
In the low temperature limit $G_2 < 0$, thus $\eta_o \le \eta \le \eta_c$. Increasing $\epsilon$ will increase the efficiency up to a critical point
$\epsilon_{crit}$ which both both $G_1 \to 0$ and $G_2 \to 0$, since $N_{h(c)}^+ \to 0$  and $N_h^- \sim N_c^-$.
At this point the engine operates at the Carnot limit and all currents vanish, such that the process becomes isoentropic,
 Cf. Fig. \ref{fig:4b}.
In the high temperature limit for harmonic oscillators, while increasing $\epsilon$, the gain $G_2$ will change sign
and become positive, thus $\eta \le \eta_o$, Cf. Fig. \ref{fig:4c}.

\subsubsection{Optimizing the amplifier's performance}
\label{subsec:Optimizing}

Further optimisation for power is obtained when
the pumping rate $\Gamma$ is optimised for maximum power then $\Gamma_{max}=2 \epsilon$ and the power becomes: 
\begin{equation}
{\cal P} = -\frac{1}{2}\hbar \nu |\epsilon| G_1~.
\label{eq:power-m}
\end{equation}

\begin{figure}
\epsfscale1200		% Figure enlarged to 120%
\center{\includegraphics[height=8cm]{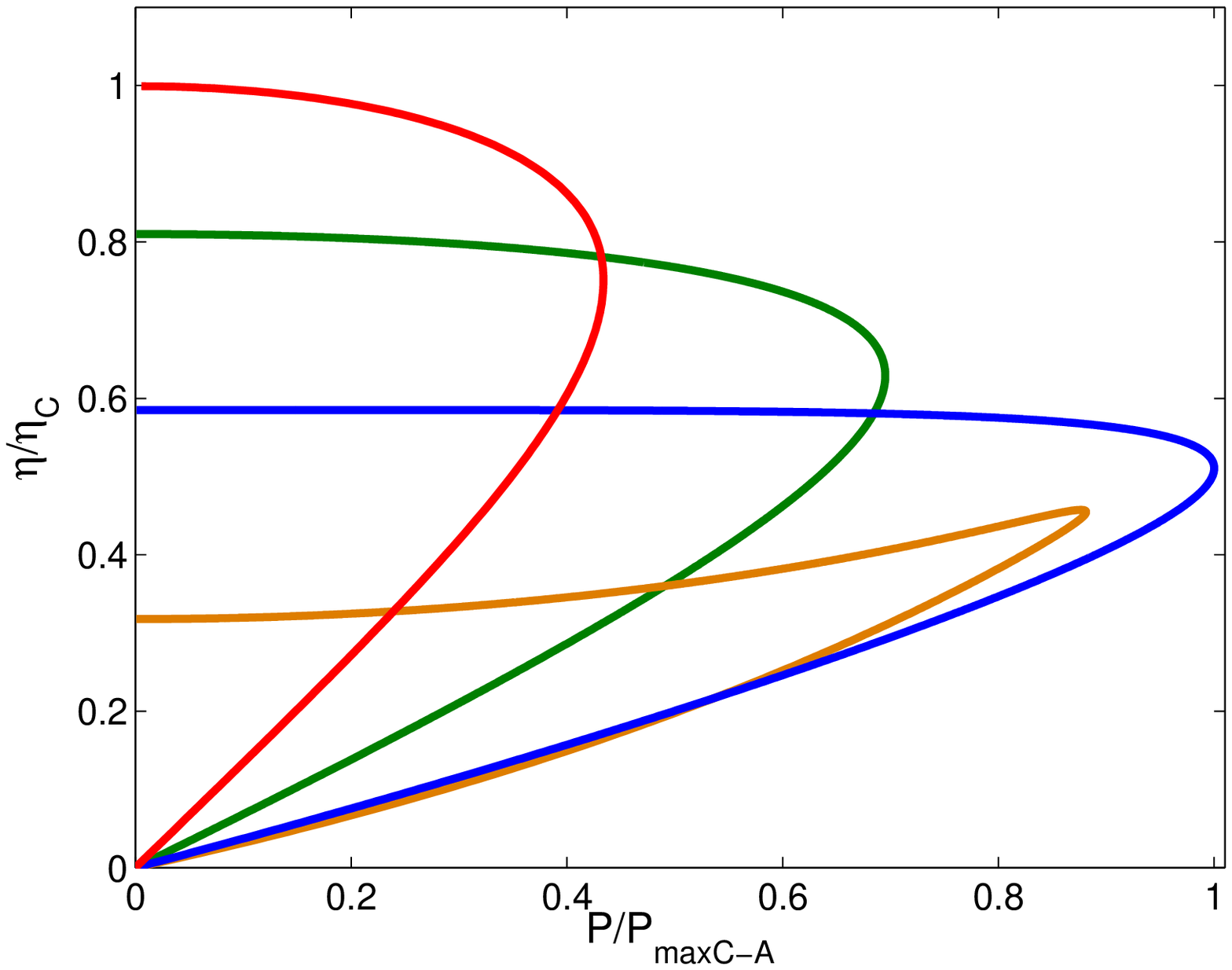}}
\caption{The normalised efficiency $\eta/\eta_c$ vs the normalised power  ${\cal P}/ {\cal P}_{max}$ for different 
optimal values of $\Gamma = 2 \epsilon$. 
${\cal P}_{max}$ is obtained for the Curzon-Ahlborn ratio $\frac{\omega_c}{\omega_h} =\sqrt{\frac{T_c}{T_h} }$
shown in blue. The orange line is for $\frac{\omega_c}{\omega_h} =\left( \frac{T_c}{T_h} \right)^{1/4}$, 
the green line is for $\frac{\omega_c}{\omega_h} =\left(\frac{T_c}{T_h} \right)^{3/4}$. The red line is 
for the Carnot ratio $\frac{\omega_c}{\omega_h} \sim \frac{T_c}{T_h} $ where the power is multiplied by $10^3$.}
\label{fig:5}
\end{figure}

At the limit of high temperature and small $\epsilon$ 
the gain becomes $G_1 \approx \frac{K_BT_h}{\hbar \omega_h}-\frac{K_BT_c}{\hbar \omega_c}$,
then optimising the power Eq. (\ref{eq:power-m}) with respect to the output frequency $\nu$ leads to: 
$$\frac{\omega_c}{\omega_h} =\sqrt{ \frac{T_c}{T_h} }~,$$ 
resulting in the efficiency at maximum power:
\begin{equation}
\eta_{CA} =1 -\sqrt{ \frac{T_c}{T_h} }~,
\end{equation}
which is the well established endoreversible result of Curzon and Ahlborn \cite{curzon75,k24}. 
The optimal power is obtained when all the characteristic parameters are balanced: $\epsilon \sim \Gamma \sim \kappa_c \sim \kappa_h$.
Figure \ref{fig:5} shows a trajectory of efficiency with respect to power with changing  field coupling 
strength $\epsilon$ for different frequency ratios $\frac{\omega_c}{\omega_h}=(\frac{T_c}{T_h})^\alpha$.
The power vanishes with the coupling $\epsilon=0$ and then when $\epsilon=\epsilon_{crit}$, at this point 
the splitting in the dressed energy levels  nulls the gain. 

%%%%%%%%%%%%%%%%%%%%%%%%%%%%%%%%%%%%%%%%%%%%%%%%%%%%%%%%%%
\subsection{Dynamical model of a 3-level engine}

The dynamical description of the 3-level engine is closely related  to the tricycle model \cite{k122}.
The model is a template for the 3-level laser shown in Fig. \ref{fig:1} with inclusion of  a dynamical description.

The Hamiltonian of the device has the form:
\begin{eqnarray}
\Op H_s = \Op H_s^0  + {\Op V(t)}=
 \left(
\begin{array}{ccc}
\epsilon_0&0&0\\
0&\epsilon_1&\epsilon e^{i \nu t}\\
0&\epsilon e^{-i\nu t}&\epsilon_2
\end{array}
\right)
\label{eq:3-level-hamil}
\end{eqnarray}
where $\Op H_s^0= \sum \epsilon_i \Op P_{ii}$, and $\Op P_{ij} = |i\rangle\langle j|$.
$\Op V(t) = \epsilon \left(\Op P_{12}e^{i\nu t}+\Op P_{21}e^{-i\nu t} \right)$.
The state of the three-level system is fully characterised by the expectation
values of any {\em eight} independent operators,
excluding the identity operator. 
Different  choices of the eight independent operators corresponds to
different viewpoints. 
\begin{itemize}
\item {\em The $P$ representation},  is based on  the eigen-representation
of the {\em free} Hamiltonian, $\Op H_s^0$, in the rotating frame.
\begin{eqnarray}
{\bf {\tilde P}}_{i,j}
= e^{-i \nu {\Op P_z} t} {\Op P_{i,j}} e^{i \nu {\Op P_z} t}~~.
\label{3l_P_rep}
\end{eqnarray}
The following notations are also used \cite{k122}:
${\bf {\hat P}_i} ~=~ {\bf {\hat P}_{ii}}$, 
${\bf {\hat P}_+} ~=~ {\bf {\hat P}_{21}}$,
${\bf {\hat P}_-} ~=~ {\bf {\hat P}_{12}}$,
${\bf {\hat P}_x} ~=~ {1 \over 2} ({\bf {\hat P}_+} + {\bf {\hat P}_-} )$,
${\bf {\hat P}_y} ~=~ {1 \over {2i}} ({\bf {\hat P}_+} - {\bf {\hat P}_-} )$,
${\bf {\hat P}_z} ~=~ {1 \over 2} ({\bf {\hat P}_u} - {\bf {\hat P}_l} )$.
\item {\em The $\Pi$ representation}, which is the eigen-representation of
the {\em full} Hamiltonian in the rotating frame,
${\bf {\tilde H}_s} = {\bf {\hat H}_s^0}  + {\bf {\tilde V}}$:
\begin{eqnarray}
{\bf {\tilde \Pi}_{i,j}} =
e^{-i \theta {\bf \tilde {P}_y}} {\bf {\tilde  P}_{i,j}}
e^{i \theta {\bf \tilde {P}_y} }~~,
\label{3l_Pi_rep}
\end{eqnarray}
where $\tan ( \theta ) = 2 \epsilon /\Delta \omega$. 
Note that
${\bf {\hat H}_s^0} + {\bf {\tilde V}}
= \Delta \omega  {\bf {\hat P}_z} + 2 \epsilon {\bf {\tilde P}_x}
= \nu {\tilde \Pi_z}$ where $\Delta \omega = \omega_h-\omega_c-\nu$.
Further notations are introduced by
${\bf {\tilde \Pi}_i}, {\bf {\tilde \Pi}_\pm}, {\bf {\tilde \Pi}_x},
{\bf {\tilde \Pi}_y}, {\bf {\tilde  \Pi}_z}$, 
defined in a similar manner to the analogous operators in the $P$ representation, provided that each $\Op P$
is replaced by a $\Op \Pi$. The $\Pi$ representation coincides with the
atom-field dressed state representation \cite{cohen87}. This
representation is employed to obtain the master equation for the engine \cite{k122}.
\end{itemize}

\begin{figure}
\epsfscale1200		% Figure enlarged to 120%
\center{\includegraphics[height=6cm]{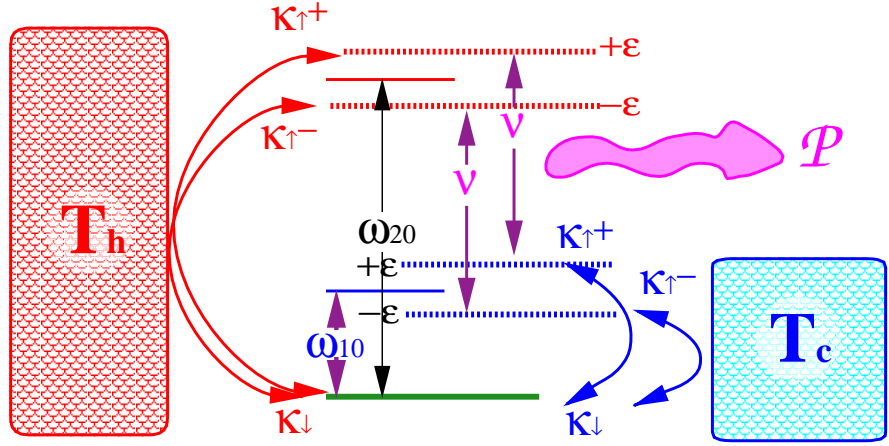}}
\caption{The three-level-amplifier: The interaction with the external field splits the two upper levels.
As a result the heat transport terms are modified. Two parallel engines emerge which can operated in
opposite direction producing zero power for a certain choice of pararmeters.}
\label{fig:6}
\end{figure}

The power and
heat fluxes only depend on a reduced subset of SU(2) operators, which are decoupled
from the rest of the operators. As a result,
the equations of motion of the 3-level amplifier can be represented by the set:
$\bf {\tilde P}_+$, $\bf {\tilde P}_-$, $\bf {\tilde  P}_u$, $\bf {\tilde P}_l$.
The final equations of motion for these observables 
and the expressions for the power and efficiency become up to numerical factors identical to the 
expressions obtained for the driven quantum tricycle. Figure \ref{fig:6} shows a schematic view of the 
splitting of the energy levels of the 3-level system due to the driving field. As a result, the engine also splits into two parts,
the upper and lower manifolds.

\subsubsection{The Generalized Lamb Equations in the $P$ Representation}
\label{subsec:gle_P_rep}

The basic equations describing the operation of a 3-level laser have been formulated by Lamb \cite{lamb64,louisell}.
These equations are modified to include the dynamical effects induced by the driving field. 
The equations of motion for the closed set of operators characterising the heat currents become:
{\small
\begin{eqnarray}
{d \over {dt}}
\left( \begin{array}{c}
{\bf \hat P_+} \\
{\bf \hat P_-} \\
{\bf \hat P_u} \\
{\bf \hat P_l}
\end{array} \right)
 = \left(
\begin{array}{cccc}
i \Delta \omega + \Gamma_+     &
0                              &
-i \epsilon + \Gamma_{+u}      &
 i \epsilon + \Gamma_{+l}      \\
0			       &
-i \Delta \omega + \Gamma_-    &
 i \epsilon + \Gamma_{-u}      &
-i \epsilon + \Gamma_{-l}      \\
-i \epsilon + \Gamma_{u+}      &
 i \epsilon + \Gamma_{u-}      &
\Gamma_u                       &
\Gamma_{ul}                    \\
 i \epsilon + \Gamma_{l+}      &
-i \epsilon + \Gamma_{l-}      &
\Gamma_{lu}                    &
\Gamma_l                       \\
\end{array}
\right)
\left( \begin{array}{c}
{\bf \hat P_+} \\
{\bf \hat P_-} \\
{\bf \hat P_u} \\
{\bf \hat P_l}
\end{array} \right)
+
\left( \begin{array}{c}
\gamma_+ \\
\gamma_- \\
\gamma_u \\
\gamma_l \\
\end{array} \right)
\label{gle_P_rep}
\end{eqnarray}
}
The relaxation coefficients $\{ \Gamma_{i,j}, \Gamma_{i}, \gamma_i \}$
consist of linear combinations of the bath parameters and can be found explicitly in Ref. \cite{k122}. 

The steady-state power and heat flows can be evaluated by
solving the generalised Lamb equations. The results are represented by Eq. (\ref{eq:flows})
where the definition of population is changed from that of a harmonic oscillator to a two-level-system: 
$N_{h(c)}^{\pm}= \frac{1}{e^{\beta_{h(c)}\omega_{h(c)}^{\pm}}+1}$.

The numerator of the expressions for the power and heat flows consist of a sum of two contributions: 
one, which is proportional to the gain $N_h^+ - N_c^+$, is 
associated with  the population inversion in the upper manifold,
while the other, which is proportional to the gain $N_h^- - N_c^-$,
is associated with the population inversion in the lower manifold. 

At low temperature the 3-level amplifier model is equivalent to that of the harmonic tricycle.
At the high temperature limit the gain $G_2$ is positive for all driving conditions $\epsilon$.
The maximum power point is the result of
competition between the upper and lower manifolds.
As $\epsilon$ {\em increases},
$N_h^+ - N_c^+$ {\em increases}, whereas $N_h^- - N_c^-$
{\em decreases}.
Hence, the power production of the upper manifold {\em increases},
(i.e. becomes more negative)
whereas that of the lower manifold {\em decreases}
(i.e. becomes less negative). 

From a thermodynamic point of view, each manifold is associated
with a separate heat engine. As the coupling with the work reservoir
($\epsilon$) increases,  the engine associated with the upper manifold
operates {\em faster} while that associated with the lower one operates
{\em slower}. In addition, the energy is leaking from the former to the latter,
thereby diminishing the net power production.

Not only does the power decrease as a function of $\epsilon$,
it also changes sign at a certain {\em finite}
value of the field amplitude,
denoted by $\epsilon_{crit}$~.
This results from the fact that at some point the lower manifold starts
operating backwards as a heat pump, rather than a heat engine.
At $\epsilon = \epsilon_{crit}$, the power {\em consumption}
by the lower manifold is exactly balanced by the power
{\em production} of the upper manifold, such that
the {\em net} power production is zero.

An examination of the steady-state heat fluxes, reveals that they
do {\em not} vanish at zero-power operating conditions. Although both manifolds 
operate at the same rate and in opposite directions, the upper manifold
absorbs more heat from the hot bath than that rejected by the lower one.
Similarly, the upper manifold rejects more heat into the cold bath than that
absorbed by the lower one. The net heat absorbed from the hot bath and
rejected into the cold bath in zero-power operating conditions is therefore
proportional to the difference in the energy gaps associated with these
transitions.
The entropy generated exclusively by the heat leak from the upper to lower manifold is obtained by eliminating
the power in Eq. (\ref{eq:pjj}):
\begin{equation}
\frac{d}{dt} \Delta {\cal S}^l ~=~ \frac{\hbar \epsilon \Gamma G_2}{4 k_B}\left(\frac{1}{T_c}-\frac{1}{T_h}\right) \ge 0~.
\label{eq:leak}
\end{equation}

Zero-power operating conditions  correspond to the short circuit limit. 
Heat is effectively transferred {\em from} the hot bath {\em into}
the cold bath, such that no net work is involved.

An interesting limit is that of low temperatures, such that
$\hbar \epsilon/k_B  \gg T_c, T_h$. This limit is realistic in
the optical domain where $\nu \gg T_c, T_h$. 
In such a case $N_h^+ - N_c^+$
is negligible relative to $N_h^- - N_c^-$, and
only the lower manifold needs be accounted for.

The value of $\epsilon_{crit}$ in this limit, denoted by $\epsilon_{crit}^{lT}$,
is that for which $N_h^- - N_c^-  =0$. It is generally given by:
\begin{eqnarray}
\epsilon_{crit}^{lT} =
{{T_h \omega_c - T_c \omega_h}
\over {T_h - T_c}}~~.
\label{e_max_lT}
\end{eqnarray}
The condition $T_h \omega_c - T_c \omega_h > 0$ is
necessary and sufficient for lasing in the limit of weak driving fields.
Population inversion becomes increasingly
more difficult in this manifold as the field intensifies.  It is therefore no longer
a sufficient condition for lasing in intense driving fields. 
Finally, zero-power operating conditions in the low-temperature limit,
obtained at $\epsilon = \epsilon_{crit}^{lT}$,
asymptotically implies zero heat flows
and hence the reversible limit.
Indeed, substituting $\epsilon_{crit}^{lT}$ from Eq. (\ref{e_max_lT})
for $\epsilon$  reduces to the Carnot efficiency,
$\eta_c=1-T_c/T_h$. 

\begin{figure}
\epsfscale1200		% Figure enlarged to 120%
\center{\includegraphics[height=8cm]{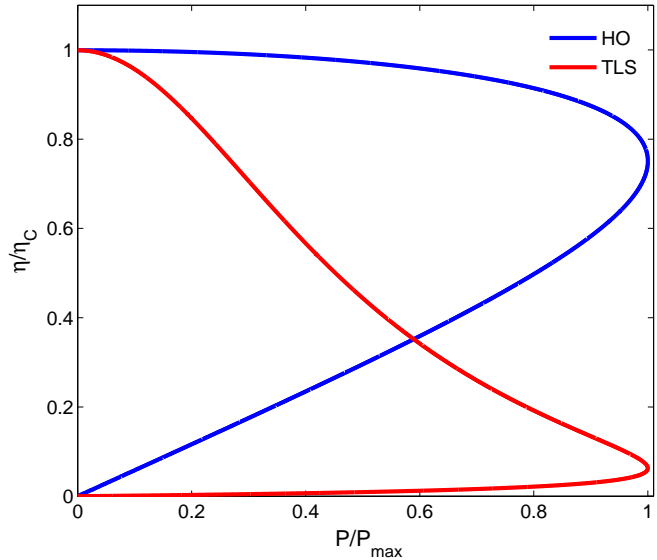}}
\caption{The normalised efficiency $\eta/\eta_c$ vs the normalised power   ${\cal P}/ {\cal P}_{max}$
for $\frac{\omega_c}{\omega_h}  \sim \frac{T_c}{T_h} $ for varying $\Gamma=2 \epsilon$. 
The high temperature limit is shown where $\hbar \omega \ll k_bT$.
The red line is the 3-level amplifier while the blue
is the harmonic  tricycle engine. Saturation limits the performance of the 3-level engine.}
\label{fig:7}
\end{figure}

The difference between a 3-level-amplifier and the tricycle  with harmonic oscillator filters at the high temperature limit
is observed in Fig. \ref{fig:7}. The source of the difference is the saturation of finite levels.
The efficiency at maximum power is lower than the Curzon-Ahlborn efficiency
approximated by:
\begin{equation}
\eta_{pmax} \approx \eta_c - \sqrt{\frac{k_B T_c}{\hbar \omega_h}}\eta_c^{\frac{1}{2}}~.
\end{equation}

\subsection{The four-level engine and two-level engine}

\begin{figure}
\epsfscale1200		% Figure enlarged to 120%
\center{\includegraphics[height=6cm]{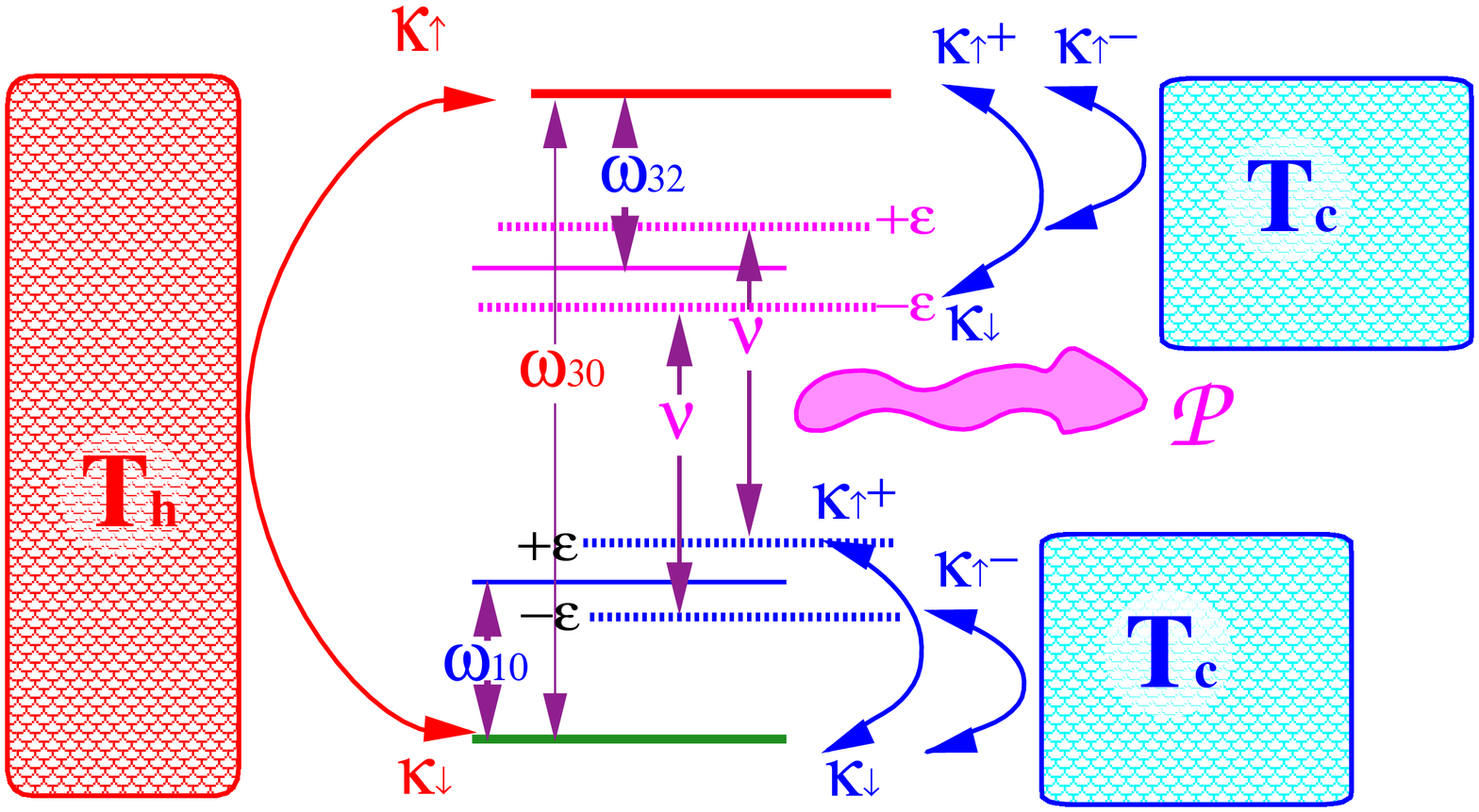}}
\caption{The four-level-engine: The interaction with the external field splits the two upper levels.
As a result the heat transport terms are modified. Two parallel engines emerge which can operated in
opposite direction producing zero power for a certain choice of pararmeters.
Coherence between levels 0 and 1 can enhance the performance by synchronising the engines.
$\omega_h=\omega_{30}$, $\omega_{c1}=\omega_{32}$ and $\omega_{c2}=\omega_{10}$.}
\label{fig:8}
\end{figure}

The four level-engine is a dynamical model of the 4-level laser \cite{digonnet90}.
In the four-level engine the pumping excitation step is isolated from the coupling to the external field by employing an additional cold reservoir.
Figure \ref{fig:8} shows a schematic view of the structure of the engine. Analysing a static viewpoint, positive gain is obtained
when 
\begin{equation}
G =p_3-p_2 =p_0\left( e^{-\frac{\hbar \omega_h}{K_B T_h}}~e^{\frac{\hbar \omega_{c2}}{k_B T_c}} -e^{-\frac{\hbar \omega_{c1}}{k_B T_c}}\right) \ge 0~,
\label{eq:flevell}
\end{equation}
where: $\omega_h=\omega_{30},~\omega_{c1}=\omega_{32},~\omega_{c2}=\omega_{10}$.
The optimal output frequency at resonance is $\nu=\omega_h-\omega_{c1}-\omega_{c2}$, therefore the efficiency becomes: 
$\eta_o = \frac{\nu}{\omega_h} $.
If the additional cold reservoir is assumed to have temperature $T_c$, then the Carnot restriction is obtained: $\eta_o \le \eta_c$.

A dynamical analysis reveals a splitting of the energy levels that are driven by the external field, leading to a similar 
performance as the three-level amplifier. The advantage of the four-level engine is that the hot bath thermal pumping 
is isolated from the coupling
to the power extraction, thus replacing the decoherence associated with the hot bath with a quieter operation
associated with the cold bath.
In addition, optimising performance by
balancing  the rates between the upper and lower manifold can be achieved by $\omega_{c1}=\omega_{c2}$.

The four level engine has been employed to study the influence of initial coherence
on the performance of the engine \cite{scully03,scully10,scully11,scully11b,rahav12,mukamel12,harbola13}.
In this case the power output is connected to the two upper levels and the coherence is generated by splitting
the ground state.
The basic idea is that initial coherence present between energy levels associated with the entangling operator $\Op Y$, 
Eq. (\ref{eq:yyy}) can be exploited to generate additional power even from a single bath. 
There is no violation of the second-law since initial coherence reduces the initial entropy.
Contact with a single bath will increase this entropy. A unitary manipulation can then exploit this entropy difference to generate work.

\begin{figure}
\epsfscale1200		% Figure enlarged to 120%
\center{\includegraphics[height=6cm]{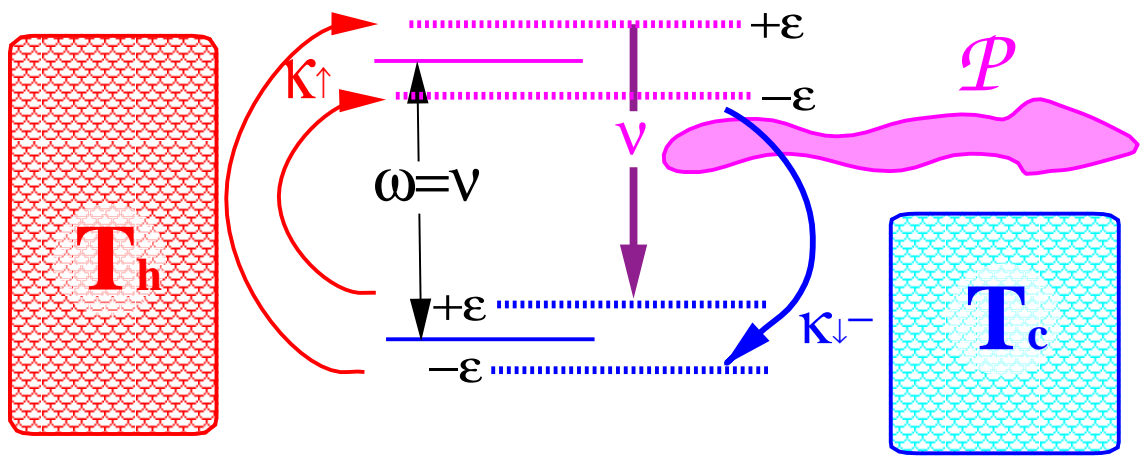}}
\caption{The two-level-engine: The interaction with the external field splits the lower and  upper levels.
If the coupling to the hot and cold bath are engineered properly output power can be produced due to the 
positive gain between the upper split levels. Only the transitions leading to this mechanism are indicated.
}
\label{fig:9}
\end{figure}

An extreme exploitation of the dynamical splitting of the energy levels due to the external driving field 
results in the two-level engine. 
Figure \ref{fig:9} sketches the schematic operation of the engine. Two excitations from the hot bath are required to
generate  a gain in the upper manifold. A clever choice of coupling to the bath is required to generate this gain.
A model of such an engine has been worked out \cite{gershon13}.
The maximum efficiency of such an engine becomes $\eta \le \frac{1}{2} \le \eta_c$ since two-pump steps
are required for one output step.

\subsection{Power storage}

An integral part of an engine is a device which allows to store the power and retrieve it on demand, a flywheel.
A model based on the tricycle of such a device just disconnects the power bath and stories the energy in the power oscillator 
$\omega_w \Op c^{\dagger} \Op c$. Once there is positive gain generated by the hot and cold bath energy will flow to this oscillator.
A model based on the 3-level amplifier suggested this approach \cite{erez07,erez08,erez13}.  The drawback of these studies is that they use
a local description of the master equation. When the flywheel oscillator has large amplitude it will modify
the system bath coupling in analogy to the dressed state picture \cite{k122}.

Other studies also considered coupling to a cavity mode \cite{rahav12,harbola13,alicki13}.
The issue to be addressed is the amount of energy that can be stored and then extracted. 
Since the storage mode is entangled to the rest of the engine this issue is delicate \cite{alicki13}.  
The amount of possible work to be extracted is defined by an entropy preserving 
unitary transformation to a passive state \cite{gelbwaser2013}.

\section{Continuous refrigerators}
\label{sec:fridge}

In a nutshell, refrigerators are just inverted heat engines. They employ power to pump heat from a cold to a hot bath.
As in heat engines, the first and second law of thermodynamics imposes the same restrictions on the refrigerator's performance. 
The destinction between the static and dynamical viewpoints also applies. 
The key element in a refrigerator is entropy extraction and disposal. 
This entropy disposal  problem is enhanced at low temperature where the entropy production in the cold bath can diverge.
 The unique feature of refrigerators is therefore  the third-law of thermodynamics.  

\subsection{The third law of thermodynamics}

Two independent formulations of the third-law of thermodynamics have been presented, both originally stated by
Nernst \cite{nerst06,nerst06b,nerst18}. The first is a purely static (equilibrium) one, also known as the "Nernst heat theorem",
phrased:
\begin{itemize}
\item{The entropy of any pure substance in thermodynamic equilibrium approaches zero as the temperature approaches
zero.}
\end{itemize}
The second formulation is  known as the unattainability principle:
\begin{itemize}
\item{It is impossible by any procedure, no matter how idealised, to reduce any assembly to absolute zero temperature in a finite number of operations \cite{Fowler39,nerst18}}.
\end{itemize}
There is an ongoing debate on the relations between the two formulations and their relation 
to the second-law regarding which and if at all, one of these formulations implies the other \cite{landsberg56,landsberg89,belgiorno03,belgiorno03b}. 
Quantum considerations can illuminate these issues. 
The tricycle model is the template used to analyse  refrigerators performance as $T_c \to 0$.

We first analyse the implications of "Nernst heat theorem" and its relations to the second-law of thermodynamics.
At steady state the second law implies that the total entropy production is non-negative, cf. Eq. (\ref{eq:II-law-t}):
$$\frac{d}{dt}\Delta S^u = -\frac{{\cal J}_c}{T_c}-\frac{{\cal J}_h}{T_h}-\frac{{\cal J}_w}{T_w}~\ge 0~~.$$
This behaviour can be quantified by a scaling exponent,
as $T_c \rightarrow 0$ the cold current should scale  with temperature as:$${\cal J}_c \propto T_c^{1+\alpha}~,$$
with an exponent $\alpha$.

When the cold bath  approaches the absolute zero temperature,
it is necessary to eliminate the entropy production divergence at the cold side,  
when $T_c \rightarrow 0$ the entropy production scales as: 
\begin{equation}
\Delta \dot S_c \sim - T_c^{\alpha}~~~,~~~~\alpha \geq 0~~.
\label{eq:III-1}
\end{equation}
For the case when $\alpha=0$ the fulfilment of the second law depends on the entropy production 
of the other baths $-\frac{{\cal J}_h}{T_h}-\frac{{\cal J}_w}{T_w} >0$, 
which should compensate for the negative entropy production of the cold bath. 
The first formulation of the third-law modifies this restriction. Instead of $\alpha \geq 0$ the third-law imposes $\alpha > 0 $ 
guaranteeing that at  absolute zero the entropy production at the cold bath is zero: $\Delta \dot S_c = 0$. 
Nernst's heat theorem then leads to the scaling condition of the heat current  with temperature \cite{k156} :
\begin{equation}
{\cal J}_c \sim T_c^{\alpha+1}~~~and~~~~\alpha > 0~.
\label{eq:III-n}
\end{equation}

We now examine the unattainability principle. 
Quantum mechanics enables a dynamical interpretation of the third-law modifying the definition to:\\
 \emph{No refrigerator can cool a system to absolute zero temperature at finite time}. 
 
This form of the third-law is more restrictive, imposing limitations on the system bath interaction and the cold bath properties when $T_c \rightarrow 0$ \cite{k275}. 
To quantify the  unattainability principle, the rate of temperature decrease of the cooling process should vanish according to the characteristic exponent $\zeta$: 
\begin{equation}
\label{eq:cp}
 \frac{dT_c(t)}{dt} = -c ~T_c^{\zeta}, ~~~ T_c\rightarrow 0 ~~.
\end{equation}
where $c$ is a positive constant. 
 Solving Eq. (\ref{eq:cp}),  leads to: 
 \begin{eqnarray}
 \begin{array}{l}
 T_c(t)^{1-\zeta}=T_c(0)^{1-\zeta}~-ct~~~~~,~for~\zeta < 1~~,\\
 T_c(t) ~=~T_c(0)e^{-ct} ~~~~~~~~~~~~~~,~for~\zeta = 1~~,\\
 \frac{1}{T_c(t)^{\zeta-1}}=\frac{1}{T_c(0)^{\zeta-1}}~~~+ct~~~~~~,~for~\zeta >1~~,
 \label{eq:iii-law-1} 
 \end{array}
 \end{eqnarray}
From Eq. (\ref{eq:iii-law-1}) it is apparent that the cold bath can be cooled to zero temperature at finite time for $\zeta < 1$. 
 The third-law requires therefore $\zeta \ge 1$. 
 The two third-law scaling relations can be related by accounting for the heat capacity  $c_V(T_c)$ of the cold bath:
\begin{equation}
 {\cal J}_c(T_c(t)) = -c_V(T_c(t))\frac{dT_c(t)}{dt}~~.
 \label{25}
\end{equation}
$c_V(T_c)$ is determined by the behaviour of the degrees of freedom of the cold bath at low temperature
where $c_V \sim T_c^{\eta}$ when $T_c \rightarrow 0$.  
Therefore the scaling exponents are related $\zeta=1+\alpha - \eta$
 \cite{k281}.

The third-law scaling relations Eq.  (\ref{25}) and Eq. (\ref{eq:III-n}) can be used as an independent check 
for quantum refrigerator models \cite{k278}. Violation of these relations points to flaws in the quantum model
of the device, typically in the derivation of the master equation.

\subsection{The quantum power driven refrigerator}

Laser cooling is a crucial technology for realising quantum devices. When the temperature is decreased,
degrees of freedom freeze out and systems reveal their quantum character. Inspired by the 
mechanism of solid state lasers and their analogy with Carnot engines \cite{scovil59} it was realised that inverting the
operation of the laser at the proper conditions will lead to refrigeration
\cite{geustic59,pikuji63,Dousmanis64,geustic68}. 
A few years later a different approach for laser cooling was initiated based on the doppler shift \cite{wineland75,hansch75}.
In this scheme, translational degrees of freedom of atoms or ions were cooled by laser light detuned to the red of the atomic transition.
Unfortunately the link to thermodynamics was forgotten. 
A flurry of activity then followed with the realisation that the 
actual temperature achieved were below the Doppler limit 
$k_B T_{doppler} =h \gamma/2 $ where $\gamma$  is the natural line width  \cite{philips88,chu89,cohen89}. 

Elaborate  quantum theories were devised to unravel the discrepancy.  The recoil limit  is based on the assumption
that the photon momentum is in quasi equilibrium with the momentum of the atoms ${\bf p }= \hbar k$. 
As a result, the temperature is related to the average kinetic energy:
\begin{equation}
k_B T_{recoil} =\frac{\hbar^2 k^2}{ 2 M}~,
\label{eq:recoil}
\end{equation}
where $\hbar k$ is the photon momentum and $M$ the atomic mass of the particle being cooled.
For sodium the doppler limit is $T_{doppler}=235 \mu K$ and the recoil limit $T_{recoil}=2.39 \mu K$ \cite{steck2003}.

Thermodynamically, these limits do not  bind, the only hard limit is the absolute zero $T_c =0$. 
To approach this limit the cooling power has to be optimised to match the cold bath temperature. 
Unoptimised refrigerators become restricted by  a minimum temperature above the absolute zero.

A reexamination of the elementary 3-level model stresses this point. Examining the heat engine model of Fig \ref{fig:1} 
shows that if the power direction is reversed  a refrigerator is generated provided the gain is negative: $G =p_2 -p_1 < 0$.
Assuming quasi equilibrium conditions. This leads to: 
\begin{equation}
\frac{\omega_c}{\omega_h} =\frac{\omega_{10}}{\omega_{20}} \le \frac{ T_c}{T_h}~,
\label{eq:ratio1}
\end{equation}
which translated to a minimum temperature of $T_c(min) \ge \frac{\omega_c}{\omega_h} T_h$.
We can consider typical values for laser cooling of Na based on the D line of $589.592~nm$ which 
translates to $\omega_h=508.838\cdot 10^{13}Hz$. If the translational cold bath is coupled to the hyperfine structure  splitting 
$F_2  \to F_1$ of the ground state $3S _{1/2}$, meaning $\omega_c =1.7716 \cdot 10^{10}Hz$.  The cooling ratio
becomes: $\frac{\omega_c}{\omega_h} = 3.48 \cdot 10^{-6}$ and the minimum temperature 
assuming room temperature of the hot bath is $T_c(min) \sim 1.5 10^{-3} K$. By employing a magnetic field
the hyperfine levels $F_1$ split in three, leading to further splittings in the MHz limit increasing the cooling ratio by 
three orders of magnitude.
When the cooling is in progress, the hyperfine splitting can be reduced by changing the magnetic 
field such that it follows the cold bath temperature. In principle any temperature above $T_c = 0$ is reachable. 

\subsection{Dynamical refrigerator models}

A quantum dynamical framework of the refrigerator is based on the tricycle template.  
The work bath is replaced by a time dependent driving term.
The cooling  current can be calculated analytically. The derivation of the equation of motion are identical to the heat engine model Sec. \ref{subsec:amplifier}.
The dynamical equations of the heat engine Eq. (\ref{eq:flows})  
converts to those of a refrigerator by inverting the sign of all heat currents. As a consequence the gain $G_1 \le 0$, Eq. (\ref{eq:gain}).
For example, when $\epsilon > \kappa$ the cooling current becomes \cite{k275}:
\begin{equation}
{\cal J}_c ~=~\frac{1}{2}\left(
\hbar \omega_c^- \frac{\Gamma_c^- \Gamma_h^-}{\Gamma_c^- +\Gamma_h^-}  (N_c^--N_h^-)
+
\hbar \omega_c^+ \frac{\Gamma_c^+ \Gamma_h^+}{\Gamma_c^+ +\Gamma_h^+}  (N_c^+ -N_h^+)
\right)~.
\label{eq:jccool}
\end{equation}
The phenomena that the refrigerator splits into two parts is also found here.
In a similar fashion the 3-level amplifier can be reversed to become a refrigerator \cite{k122}. 
The heat currents are identical to Eq. (\ref{eq:flows}) with the TLS definition of $N_{c/h}^{\pm}$.

In a refrigerator the object of optimisation is the  cooling power ${\cal J}_c$ compared to the output power ${\cal P}$ in an engine.
The efficiency is defined by the coefficient of performance COP the ratio between ${\cal J}_c$ and the input power ${\cal P}$. 
\begin{equation}
COP = \frac{{\cal J}_c}{\cal P} \le COP_o \le COP_c
\label{eq:cop}
\end{equation}
where $COP_o=\frac{\omega_c}{\nu} $ and $COP_c=\frac{T_c}{T_h-T_c}$.
All power driven refrigerators are restricted by the Otto $COP_o$. 
In section \ref{subsec:elawr} the performance of power driven refrigerators at low temperature
will be analysed.

\subsection{The quantum absorption refrigerator}

The absorption chiller is a refrigerator which employs a heat source to replace mechanical work for
driving a heat pump \cite{jeff00}.  The first device was developed in 1850 by the Carr\'e brothers  which became the first useful refrigerator.
In 1926 Einstein and Szil\'ard invented an absorption refrigerator with no moving parts \cite{szilard1926}. 
This fact is an inspiration for miniaturizing the device to solve the growing problem of heat generated in
integrated circuits. Even a more challenging proposal is to miniaturize to the level of 
a few-level quantum system. Such a device could be  incorporated into a quantum circuit. 
The first quantum version was based on the 3-level refrigerator \cite{k169} driven by thermal noise.
A more recent model of an autonomous quantum absorption refrigerator with no external intervention based 
on three-qubits \cite{popescu10} has renewed interest in such devices. 
A setup of opto-mechanical refrigerators powered by incoherent thermal light was introduced by \cite{mari2012}.
The study showed that cooling increases while increasing the photon number up to the point that fluctuation
of the radiation pressure becomes dominant and heats the mechanical mode. Studies of cooling based on
electron tunneling were also considered. The working medium is composed of quantum dots
\cite{vandebrock12,givonati2013} or NIS junction \cite{pekola12} (Normal metal Insulator Superconductor). 
Heat is removed from the reservoir by electron transport
which is induced by thermal or shot noise.

\subsubsection{The 3-level absorption refrigerator}

In the 3-level absorption refrigerator the power source has a finite temperature $T_w$. The 
coefficient of performance becomes $COP = \frac{ {\cal J}_c}{{\cal J}_w}$. From the second and first law 
Eqs. (\ref{eq:I-law-t}), (\ref{eq:II-law-t}) using steady state conditions one obtains:
\begin{equation}
COP \le \frac{(T_w -T_h)T_c}{(T_h-T_c)T_w}
\label{eq:cop-1}
\end{equation}
and $T_w > T_h > T_c$ \cite{jeff00}. Under the assumption that the three baths are uncorrelated,
the 3-level state is diagonal in the energy representation and determined by the heat conductivities \cite{k169,palao13b}.
From these values the heat currents can be evaluated. This refrigerator emphasises the fact that no internal coherence
is required for operation.

\subsubsection{Tricycle absorption refrigerator  }

The template to understand the absorption refrigerator is the tricycle model, a refrigerator connected to
three reservoirs with the Hamiltonian Eq. (\ref{eq:hfilter}) and Eq. (\ref{eq:hinteraction}). 
A thermodynamical consistent description requires to first diagonalize the Hamiltonian and then
to obtain the generalized master equations for each reservoir. The final step is to obtain the steady state
heat currents ${\cal J}_c$, ${\cal J}_h$ and ${\cal J}_w$. The nonlinearity in Eq. (\ref{eq:hinteraction}) 
hampers this task. 

\begin{figure}
\epsfscale1200		% Figure enlarged to 120%
\center{\includegraphics[height=8cm]{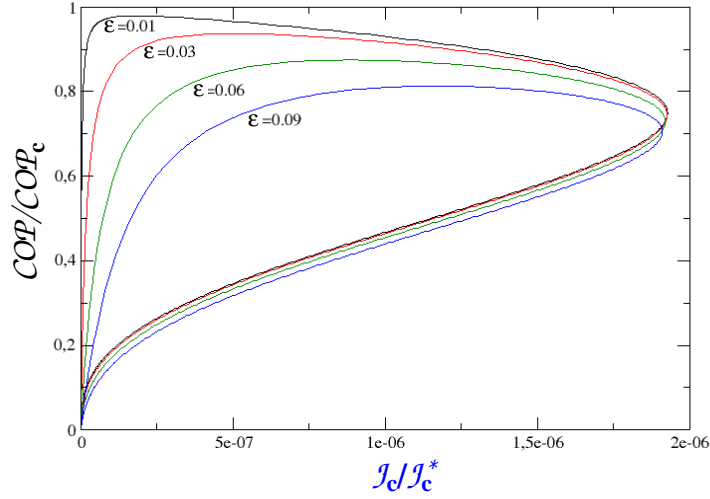}}
\caption{The $COP/COP_{c}$  of the 3-qubit refrigerator vs the normalised cold current ${\cal J}_c/{\cal J}_c^*$
for different values of the internal coupling constant $\epsilon$. The $COP$  at maximum power
is found to be bound by $3/4COP_c$ independent of $\epsilon$.
Figure in courtesy of L.A. Correa and J.P. Palao.
}
\label{fig:10}
\end{figure}

One remedy to this issue is to replace the harmonic oscillators in the model by qubits \cite{palao13}.
The other approach is to consider the high temperature limit of the power reservoir or  a noise driven refrigerator \cite{k272,k275}.
This three qubit refrigerator model  was introduced as the ultimate miniaturization model \cite{popescu10,popescu11,popescu12}.
The free Hamiltonian of the device has the form:
\begin{equation}
\Op H_F = \hbar \omega_h \Op \sigma_z^h+\hbar \omega_c \Op \sigma_z^c+ \hbar \nu \Op \sigma_z^w
\label{eq:qtricycle}
\end{equation}
and the interaction Hamiltonian: 
\begin{equation}
\Op H_I = \hbar \epsilon \left(\Op  \sigma_+^h \otimes \Op \sigma_-^c \otimes \Op \sigma_-^w + 
\Op \sigma_-^h \otimes \Op  \sigma_+^c \otimes \Op \sigma_+^w \right)~,
\label{eq:intqtricycle}
\end{equation}
where $\Op \sigma$ are two-level operators.
Assuming $\omega_h = \omega_c+\nu$
the eigenstates of the Hamiltonian $\Op H =\Op H_F+\Op H_I$ become:
\begin{eqnarray}
\begin{array}{lll}
Number&Energy&state\\
1&0&|0,0,0\rangle\\
2&\hbar \omega_c&|0,1,0\rangle\\
3&\hbar \nu&|0,0,1\rangle\\
4&\hbar(\nu+\epsilon)&\frac{1}{\sqrt{2}}(|0,1,1\rangle+|1,0,0\rangle)\\
5&\hbar(\nu-\epsilon)&\frac{1}{\sqrt{2}}(|0,1,1\rangle-|1,0,0\rangle)\\
6&\hbar(\omega_c+\omega_h)&|1,1,0\rangle\\
7&\hbar(\omega_h+\nu)&|1,0,1\rangle\\
8&2 \hbar \omega_h&|1,1,1\rangle\\
\end{array}
\label{eq:eig3bit}
\end{eqnarray}
The coupling to the bath has the form:
\begin{equation}
\Op H_{SR} =\sum_{j=c,h} \lambda_j (\Op \sigma_+^j \otimes \Op B^j_- +\Op \sigma_-^j \otimes \Op B^j_+) ~,
\label{eq:bath}
\end{equation}
where $j$ is the bath index. In this model, the coupling to the cold bath connects levels $1 \rightarrow 2~~$,
$3 \rightarrow 4,5~~$ and  $4,5 \rightarrow 8$. The master
equation for the refrigerator can be derived using the weak system bath coupling limit \cite{palao13}:
\begin{equation}
{\cal L} =\frac{i}{\hbar}[\Op H, \bullet ] +{\cal L}_h+{\cal L}_c+{\cal L}_w~.
\end{equation}
The cooling and power currents  can be solved for this model. 
Of major importance is the distinction between the complete non-local description of the tricycle
and a local description.  In the local case the dissipative terms ${\cal L}_{h/c}$ are set to equilibrate only the local qubit.
In the nonlocal case the local qubit is mixed with the other qubits due to the nonlinear coupling.
As a result in the nonlocal approach it is impossible to reach the Carnot $COP_c$ no matter how small is 
the internal coupling $\epsilon$. This can be observed in Fig. \ref{fig:10}, showing the 
$COP$ vs the cooling current ${\cal J}_c$ for different coupling strength $\epsilon$.
On the other hand the Carnot limit $COP_c$ is reachable in the local model.
An interesting observation is that a universal limit for
the $COP$ at the maximum cooling power was found \cite{palao13b} applicable
for all absorption refrigerator modes of  $COP^*  \le \frac{d}{d+1} COP_c$ where $d$ is the diemsionality
of the phonon cold bath.

\subsubsection{Noise driven refrigerator }

An ideal power source generates zero entropy: $\Delta S_w=-\frac {{\cal J}_w}{T_w}$. A pure mechanical external field
achieves this goal. Another possibility is a thermal source, when $ T_w \rightarrow \infty $ it also generates zero entropy.
An unusual power source is noise. Gaussian white noise also carries with it zero entropy production. 
Analysis will show that
the performance of refrigerators driven by all these power sources are very similar even when 
approaching the absolute zero temperature \cite{k272}.

Employing the tricycle as a template we examine the option of employing noise as a power source.
The simplest option is a Gaussian white noise source.
As a result the interaction nonlinear term Eq. (\ref{eq:hinteraction}) is replaced with:
\begin{eqnarray}
\begin{array}{rcl}
\Op H_{int}&=&  f(t)\left( \Op a^{\dagger} \Op b + \Op a \Op b^{\dagger}   \right)=  f(t) \Op X ~,
\end{array}
\label{eq:hamil2}
\end{eqnarray}
where $f (t) $ is the noise field. $\Op X=(\Op a^{\dagger} \Op b  + \Op a \Op b^{\dagger})$ is
the generator of a swap operation between the two oscillators and  is part of  a set of $SU(2)$ operators ,
$\Op Y=i(\Op a^{\dagger} \Op b  - \Op a \Op b^{\dagger})$,
$\Op Z = \left( \Op a^{\dagger}  \Op a  -   \Op b^{\dagger} \Op b \right)$ and the Casimir
$\Op N = \left( \Op a^{\dagger}  \Op a  +   \Op b^{\dagger} \Op b \right)$.

A Gaussian source of white noise is characterized by
zero mean $\langle f(t) \rangle=0$ and delta time correlation $\langle f(t) f(t') \rangle = 2 \eta \delta(t-t')$. 
The Heisenberg  equation for a time independent operator $\Op O$ reduced to: 
\begin{equation}
\frac{d}{dt} \Op O ~~=~~ i [ \Op H_s, \Op O ]  +{\cal L}_n^*(\Op O)+{\cal L}_h^* (\Op O)+ {\cal L}_c^* (\Op O)~,\label{eq:glvn}
\end{equation}
where $\Op H_s = \hbar \omega_h \Op a^{\dagger} \Op a +\hbar \omega_c  \Op b^{\dagger}\Op b $. 
The noise dissipator for Gaussian noise is ${\cal L}_n^*(\Op O) = -\eta [ \Op X, [\Op X , \Op O]]$ \cite{gorini76}. 
 
The next step is to derive the quantum Master equation of each reservoir. 
It is assumed that the  reservoirs are uncorrelated and also
uncorrelated with the driving noise. These conditions simplify the derivation of  ${\cal L}_h$ 
which become the standard energy relaxation terms, driving oscillator 
$\hbar  \omega_h \Op a^{\dagger} \Op a$ to thermal equilibrium with temperature $T_h$ and ${\cal L}_c$ 
drives oscillator $ \hbar \omega_c \Op b^{\dagger} \Op b$ to equilibrium $T_c$ \cite{breuer}.
\begin{eqnarray}
\begin{array}{rcr}
{\cal L}_{h}^* (\Op O )&~=~& \Gamma_h (N_{h} +1 ) \left( \Op a^{\dagger} \Op O  \Op a -\frac{1}{2} \left\{\Op a^{\dagger}   \Op a, \Op O \right\} \right)\\
&&~+~\Gamma_h N_{h} \left( \Op a \Op O  \Op a^{\dagger}  -\frac{1}{2} \left\{\Op a  \Op a^{\dagger} , \Op O \right\} \right)\\
{\cal L}_c^* (\Op O )&~=~& \Gamma_c (N_c +1 ) \left( \Op b^{\dagger} \Op O  \Op b -\frac{1}{2} \left\{\Op b^{\dagger}   \Op b, \Op O \right\} \right)\\
&&~+~\Gamma_c N_c \left( \Op b \Op O  \Op b^{\dagger}  -\frac{1}{2} \left\{\Op b  \Op b^{\dagger} , \Op O \right\} \right)\\
\end{array}~.
\label{eq:relaxabsor}
\end{eqnarray}
The kinetic coefficients $\Gamma_{h/c}$ are 
determined from the system bath coupling and the spectral function \cite{k122,k275}.

The equations of motion including the dissipative part are closed to the $SU(2)$ set of operators. 
To derive the cooling current ${\cal J}_c= \langle {\cal L}_c( \hbar \omega_c \Op b^{\dagger} \Op b)\rangle$,
we solve for stationary solutions of $\Op N$ and $\Op Z$.  The cooling current becomes:
\begin{eqnarray}
\begin{array}{rcl}
{\cal J}_c &~=~&  \hbar \omega_c\frac{2 \eta \bar \Gamma }{2 \eta+ \bar \Gamma}(N_c-N_h)
\end{array}~.
\label{eq:Jc}
\end{eqnarray}
where the effective heat conductance is 
$\bar \Gamma = \frac{\Gamma_c \Gamma_h}{\Gamma_c + \Gamma_h}$.
Cooling occurs for $N_c > N_h \Rightarrow \frac{\omega_h}{T_h} > \frac{\omega_c}{T_c}$. 
The coefficient of performance ($COP$) for the absorption chiller is defined by the relation 
$COP = \frac{{\cal J}_c}{{\cal J}_n}$, 
with the help of Eq. (\ref{eq:Jc}) we obtain the Otto  $COP_o$ \cite{jahnkemahler08}, Cf. Eq. (\ref{eq:cop}).

We now show the equivalence of the noise driven refrigerator with 
the high temperature limit of the work bath $T_w$. Based on the weak coupling limit the dissipative generator of 
the power bath becomes:
\begin{eqnarray}
\begin{array}{rcr}
{\cal L}_w^* (\Op O )&~=~& \Gamma_w (N_w +1 ) \left( \Op a^{\dagger} \Op b \Op O  \Op b^{\dagger} \Op a 
-\frac{1}{2} \left\{\Op a^{\dagger}   \Op a \Op b \Op b^{\dagger} , \Op O \right\} \right)\\
&&~+~\Gamma_w N_w \left( \Op a \Op b^{\dagger} \Op O  \Op a^{\dagger} \Op b 
-\frac{1}{2} \left\{\Op a  \Op a^{\dagger} \Op b^{\dagger} \Op b , \Op O \right\} \right)\\
\end{array}~.
\label{eq:relaxw}
\end{eqnarray}
where $N_w = (\exp(\frac {\hbar \omega_w}{k T_h})-1)^{-1} $. At finite temperature ${\cal L}_w(\Op O)$ does not lead to a close set of equations. But in the limit of $T_w \rightarrow \infty$ it becomes equivalent to the Gaussian noise generator: 
${\cal L}_w^* (\Op O)= -\eta/2 \left( [ \Op X , [\Op X, \Op O]]+ [ \Op Y , [\Op Y, \Op O]] \right)$, where $\eta= \Gamma_w N_w$.
This noise generator leads to the same current ${\cal J}_c$ and $COP$ as Eq. (\ref{eq:Jc}) and (\ref{eq:cop}).
We conclude that Gaussian noise represents the singular bath limit equivalent to $T_w \rightarrow \infty$. 

Poisson white noise can be employed as a power source. This noise is typically generated by a sequence of independent random pulses with exponential inter-arrival times \cite{luczka91,alicki06}. These impulses drive the coupling between the oscillators in contact with  the hot and cold bath leading to:
\begin{eqnarray}
\label{eq:master-eq}
\begin{array}{rcl}
\frac{d \Op O}{dt}&=&(i/\hbar)[\Opt H,\Op O] - (i/\hbar)\lambda \langle \xi \rangle [\Op X,\Op O] \\
 &&+\lambda\left( \int^{\infty}_{-\infty}d\xi P(\xi)e^{(i/\hbar)\xi \Op X}\Op O e^{(-i/\hbar)\xi \Op X} -\Op O \right)~,
 \end{array}
\end{eqnarray}
where $ \Opt H $ is the total Hamiltonian including the baths. $\lambda$ is the rate of events and $\xi$ 
is the impulse strength averaged over a distribution $P(\xi)$.
Using the Hadamard lemma and the fact that the operators form a closed $SU(2)$ algebra, we can separate the noise contribution to its unitary and dissipation parts, leading to the master equation,
\begin{equation}
\label{eq:vn}
\frac{d \Op O}{dt}=(i/\hbar)[\Opt H,\Op O]+(i/\hbar)[\Op H^{\prime} ,\Op O]+ {\cal L}_n^*(\Op O)~.
\end{equation} 
The unitary part is generated with the addition of the Hamiltonian
$ \Op H^{\prime}= \hbar\epsilon \Op X $ and with the interaction
\begin{equation}
\epsilon= -\frac{\lambda}{2}\int d\xi P(\xi)(2\xi/\hbar -sin(2\xi /\hbar))\nonumber~.
\end{equation}
This term can cause a direct heat leak from the hot to the cold bath.
The noise generator ${\cal L}_n(\Op\rho)$,  can be reduced to the form 
$
{\cal L}_n^*(\Op O )= -\eta [\Op X,[\Op X,\Op O]]~,
$
with a modified noise parameter:
\begin{equation}
\eta=\frac{\lambda}{4}\left( 1-\int d\xi P(\xi)cos(2\xi /\hbar)\right) \nonumber~.
\end{equation}
The Poisson noise generates an effective Hamiltonian which is composed of 
$\Opt H$ and $\Op H^{\prime}$, modifying the energy levels of the working medium.
This  new Hamiltonian structure has to be incorporated in the derivation of the master equation 
otherwise the second law will be violated. 
The first step is to rewrite the system Hamiltonian in its dressed form.
A new set of bosonic operators is defined
\begin{eqnarray}
\begin{array}{l}
 \Op A_{1} =  \Op a \cos(\theta) +\Op b \sin(\theta)  \\
 \Op A_{2} =  \Op b \cos(\theta) -\Op a \sin(\theta) ~,
 \end{array}
\end{eqnarray}
The dressed Hamiltonian is given by:
\begin{equation}
\label {eq:dressed H}
 \Op H_{s} = \hbar\Omega_{+}\Op A^{\dagger}_1 \Op A_1 + \hbar\Omega_{-}\Op A^{\dagger}_2 \Op A_2~,
\end{equation}
where
$
\Omega_{\pm} = \frac{\omega_h +\omega_c}{2} \pm  \sqrt{(\frac{\omega_h -\omega_c}{2})^2 +\epsilon^2} 
$
and
$
\cos^2(\theta) = \frac{\omega_h-\Omega_{-}}{\Omega_{+}-\Omega_{-}} 
$
Eq.(\ref {eq:dressed H}) impose the restriction, $\Omega_{\pm}>0$ which can be translated to $\omega_h \omega_c > \epsilon^2$. 
The master equation in the Heisenberg representation becomes:
\begin{equation}
\frac{d \Op O}{dt}=(i/\hbar)[\Op H_s ,\Op O] + {\cal L}_h^*(\Op O) +{\cal L}_c^*(\Op O)+{\cal L}_n^*(\Op O)~,
\end{equation}
where the details can be found in ref \cite{k275}.
The noise generator becomes:
\begin{equation}
{\cal L}_n^*(\Op O)= -\eta [\Op W,[\Op W,\Op O]]~,
\end{equation}
where $\Op W=\sin(2 \theta)\Op Z+\cos(2 \theta)\Op X$.
Again, the  $SU(2)$ algebra is employed to define the operators: 
$\Op X=(\Op A_1^{\dagger}\Op A_2+\Op A_2^{\dagger}\Op A_1)$ , 
$\Op Y=i(\Op A_1^{\dagger}\Op A_2 - \Op A_2^{\dagger}\Op A_1)$  and $\Op Z=(\Op A_1^{\dagger}\Op A_1-\Op A_2^{\dagger}\Op A_2)$. The total number of excitations is accounted for by the operator $\Op N=(\Op A_1^{\dagger}\Op A_1+\Op A_2^{\dagger}\Op A_2)$.  \\ 

Once the set of linear equations is solved the exact expression for the heat currents is extracted,
${\cal J}_h=\left\langle {\cal L}_h^*(\Op H_{s})\right\rangle $, $
{\cal J}_c=\left\langle {\cal L}_c^*(\Op H_{s})\right\rangle $ and $
{\cal J}_n=\left\langle {\cal L}_n^*(\Op H_{s})\right\rangle $. 

\begin{figure}
\epsfscale1200		% Figure enlarged to 120%
\center{\includegraphics[height=6cm]{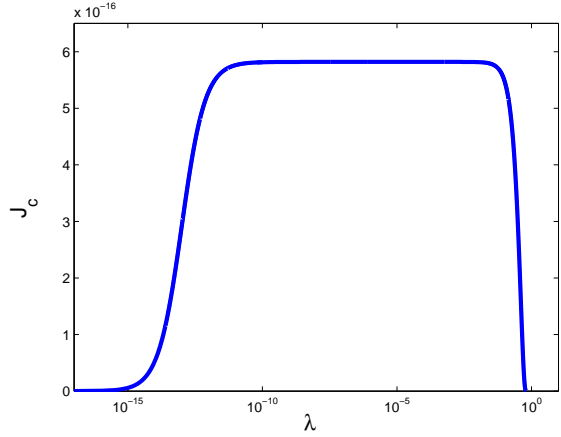}}
\caption{The noise driven refregerator. The cooling current ${\cal J}_c$ as a function of $\lambda$. 
The refrigerator is powered by
a Poissonian noise source. The plot is  for the limit of $\sigma \to 0$.  }
\label{fig:ppp}
\end{figure}

The distribution of impulse determines the performance of the refrigerator.
For a normal distribution of impulses in Eq. (\ref{eq:master-eq}),  
$P(\xi)=\frac{1}{\sqrt{2 \pi \sigma^2} }e^{-\frac{ (\xi-\xi_0)^2}{2 \sigma^2}}$.
The energy shift is controlled by:
\begin{equation}
\epsilon=-\frac{\lambda}{2}(2\xi_0 /\hbar- e^{-\frac{2 \sigma^2}{\hbar^2}}sin(2\xi_0/\hbar))~.
\label{eq:eps}
\end{equation}
The effective noise strength becomes:
\begin{equation}
\eta=\frac{\lambda}{4}(1-e^{-\frac{2 \sigma^2}{\hbar^2}}cos(2\xi_0/\hbar))~. 
\label{eq:eta-2}
\end{equation}
In the limit of $\sigma \to 0$,  $P(\xi)=\delta(\xi-\xi_0)$.
Another possibility is an  exponential distribution: $P(\xi) = \frac{1}{\xi_0}e^{-\frac{\xi}{\xi_0}}$ then:
\begin{equation}
\epsilon=-\lambda \frac{(\xi_0/{\hbar})^3}{ 4 ( 1+(\xi_0/\hbar)^2 )}
\label{eq:eps-e}
\end{equation}
and
\begin{equation}
\eta=\lambda \frac{(\xi_0/{\hbar})^2}{   1+4 (\xi_0/\hbar)^2 }~.
\label{eq:eta-e}
\end{equation}
The Poissonian noise plays two opposing roles. On the one hand it increases the cooling current ${\cal J}_c$,
by increasing $\eta$. On the other hand it decreases $\epsilon$ (becomes more negative) and by that decreases ${\cal J}_c$.
Both parameters $\eta$ and $\epsilon$ depend linearly on $\lambda$, which can be interpreted as the rate
of photon absorption in the system which enhances the cooling process.
Fig. \ref{fig:ppp} shows that ${\cal J}_c$ increases with $\lambda$ until a point where $\epsilon $ dominates
and ${\cal J}_c$ decreases.

The $COP$ for the Poisson driven refrigerator is restricted by the Otto and Carnot $COP$:
\begin{equation}
COP =\frac{\Omega_-}{\Omega_+ - \Omega_-} \le \frac{\omega_c}{\omega_h-\omega_c} \le \frac{T_c}{T_h-T_c}~.
\label{eq:cop2}
\end{equation}

\subsection{Refrigerators operating close to the limit $T_c \to 0$ and the third-law of thermodynamics}
\label{subsec:elawr}

The performance of all types of refrigerators at low temperature have  universal properties.  \\
In the power driven refrigerators  the lower split manifold is dominant, leading to the cold current:
\begin{equation}
{\cal J}_c \approx \hbar \omega_c^- \frac{2\epsilon^2 \bar \Gamma }{4 \epsilon^2+\Gamma_c \Gamma_h}\cdot G~,
\label{eq:pjc}
\end{equation}
where the gain $G=N_c^--N_h^-$ and $\bar \Gamma =\frac{\Gamma_c\Gamma_h}{\Gamma_c+\Gamma_h}$.\\
In the 3-level absorption refrigerator the cold current becomes \cite{palao13b}:
\begin{equation}
{\cal J}_c ~=~  \hbar \omega_c \frac{\Gamma_c\Gamma_h\Gamma_w}{\Delta} \cdot G
\label{eq:jose}
\end{equation} 
where $\Delta$ is a combination of kinetic constants and 
$G= e^{-\frac{\hbar \omega_w}{k_B T_w}}e^{-\frac{\hbar \omega_c}{k_B T_c}}-e^{-\frac{\hbar \omega_h}{k_B T_h}}$.\\
In the Guassian noise driven  refrigerator Eq. (\ref{eq:Jc}) becomes:
\begin{equation}
\begin{array}{rcl}
{\cal J}_c ~=~  \hbar \omega_c\frac{2 \eta \bar \Gamma  }{2 \eta +\bar \Gamma}\cdot G
\end{array}~.
\label{eq:Jc2}
\end{equation}
where $G=N_c-N_h$.\\
 In the Poisson driven refrigerator we obtain:
\begin{equation}
{\cal J}_c \approx \hbar \Omega_- \frac{2\eta \bar \Gamma  }{2\eta+\bar \Gamma}\cdot G~,
\label{eq:pjc}
\end{equation}
with $G=(N_c^- -N_h^+)$ and 
$\Omega_- \approx \omega_c -\frac{\epsilon^2}{\omega_h-\omega_c}$. 
All expressions for the cooling current ${\cal J}_c$ are a product of an energy quant $\hbar \omega_c$, 
the effective heat conductance and the gain $G$.

To approach the absolute zero temperature a further optimisation is required. 
Optimising the gain $G$ is obtained when  $\omega_h \to \infty$, which leads to $G \sim e^{-\frac{\hbar \omega_c}{k_B T_c}}$. 
In addition the driving amplitude should be balanced with the heat conductivity.
The cooling current can also be expressed in terms of the relaxation rate to the bath via the relation
$\gamma(\omega) e^{-\frac{\hbar \omega}{k_B T}}=\Gamma(\omega)  N(\omega)$.
The universal optimised cooling current as $T_c \to 0$ becomes:
\begin{equation}
{\cal J}_c ~=~ \hbar \omega_c \cdot \gamma_c \cdot e^{-\frac{\hbar \omega_c}{k_B T_c}}~.
\label{eq:3law1}
\end{equation}
This form is  correct both in the weak coupling limit and the  low density limit. The heat current ${\cal J}_c$
can be interpreted as the quant  of energy $\hbar \omega_c$ extracted from the cold bath
at the rate $\gamma_c$ multiplied by a Boltzmann factor. 
Further optimisation with respect to $\omega_c$ is dominated by the
exponential Boltzmann factor (Optimising the function $x^a e^{-x/b} $ leads to $x^*\propto b$).
As a result $\omega_c^* \propto T_c$, obtaining: ${\cal J}_c \propto \omega_c^* \cdot \gamma_c(\omega_c^*)$.
The linear relation between the optimal frequency $\omega_c$ and $T_c$ allows to translate the temperature scaling relations
to the low frequency scaling relations of $\gamma_c(\omega ) \sim \omega^{\mu}$ and $c_V(\omega) \sim \omega^{\eta}$ when $\omega \to 0$.

To fulfil "Nernst's heat theorem" Eq. (\ref{eq:III-n}), the scaling of the relaxation rate is restricted to
$\gamma_c (\omega) \sim \omega^{\alpha}$ and $\alpha > 0$.
The fulfilment of the unattainability principle Eq. (\ref{25}) depends on the ratio between the relaxation rate and the heat capacity 
$\gamma_c/c_V \sim \omega^{\zeta-1}$ where $\zeta > 1$. 

The  low frequency properties of
the relaxation rate $\gamma_c$ and the heat capacity $c_V$ are examined. 
For three-dimentional ideal degenerate Bose  gases $c_V$ scales as $T_c^{3/2}$. For degenerate Fermi gas $c_V$ scales as $T_c$.
In both cases the fraction of the gas that can be cooled decreases with temperature. 
Based on a collision model when cooling occurs due to inelastic scattering we have found the scaling exponent
$\zeta=3/2$ for both Bose and Fermi  degenerate gas \cite{k275}.

The most studied generic bath is the harmonic bath. 
It includes the electromagnetic field: (a photon bath), or a macroscopic piece of solid;
(a phonon bath), or Bogoliubov excitations in a Bose-Einstein condensate.
The standard form of the bath's Hamiltonian is:
\begin{equation}
\Op H_{int} = (\Op b + \Op b^{\dagger})\left(\sum_{k} (g(k)\Op a(k) + \bar{g}(k)\Op a^{\dagger}(k))\right)\ ,
\ \Op H_B = \sum_{k}\omega(k)\Op a^{\dagger}(k)\Op a(k)~~
\label{26}
\end{equation}
where $\Op a(k)$ and $\Op a^{\dagger}(k)$ are the annihilation and creation operators for a mode $k$.
For this model, the weak coupling limit procedure leads to the GKS-L generator with 
the cold bath relaxation rate given by \cite{k275}:
\begin{equation}
\gamma_c \equiv \gamma_c(\omega_c) = \pi~\sum_{k} |g(k)|^2 \delta (\omega(k) - \omega_c)\left[1- e^{-\frac{\hbar \omega(k)}{k_BT_c}}\right]^{-1}~~.
\label{27}
\end{equation}
For the bosonic field in $d$-dimensional space, with the linear low-frequency dispersion law ($\omega(k) \sim |k|$),
the following scaling properties for the cooling rate at low frequencies are~obtained:
$
\gamma_c \sim \omega_c^{\kappa}\omega_c^{d-1} 
$
where $\omega_c^{\kappa}$ represents the scaling of  the form-factor $|g(\omega)|^2$, and $\omega_c^{d-1}$
is the scaling of the density of modes. For $\omega_c \sim T_c$,  the final current scaling becomes:
${\cal J}_c^{opt} \sim T_c^{d+\kappa}$ or $\alpha =d +\kappa-1$.

At low temperatures, the heat capacity of the bosonic systems scales like:
$c_{V}(T_c) \sim T_c^d$,
which produces the scaling $\zeta = \kappa$. This means that to fulfil the third law $ \kappa \geq 1 $.
To rationalise the scaling,  $c_V$ is a volume property and $\gamma_c$ is a surface property,
so  $\omega_c \gamma_c$ scales the same as $c_V$. The scaling of the form factor $\kappa$
is related to the speed the excitation can be carried away from the surface.

The exponent $\kappa =1$ when $\omega \to 0$,
is typical in systems such as electromagnetic fields or acoustic phonons  which have  linear dispersion law, $\omega(k)= v|k|$. 
In these cases the form-factor becomes, $g(k)\sim |k|/\sqrt{\omega(k)}$, therefore $|g(\omega)|^2 \sim |k|$. 
The condition $ \kappa \geq 1 $ excludes exotic
dispersion laws, $\omega(k)\sim |k|^{\delta}$ with $\delta < 1$, which produce infinite group velocity forbidden by relativity theory. 
Moreover, the popular choice of Ohmic coupling is excluded for systems in dimension $d > 1$ \cite{k281}. 
We should mention that challenges to the third law have been published \cite{gershon12}.
Our viewpoint is  that these discrepancies are the result of flaws in reduction of the spin bath 
to a harmonic spectral density formulation.

\section{Summary}

Thermodynamics represents physical reality with an amazingly small number of variables. 
In equilibrium, the energy operator $\Op H$ is sufficient to reconstruct the state of the system, from which
all other observables can be deduced.
Dynamical systems out of equilibrium require more variables. 
Quantum thermodynamics advocates that only a few additional variables are required to  describe
a quantum device; a heat engine or a refrigerator.
A description based on Heisenberg equation of motion lends itself to this viewpoint. 
A sufficient condition for the set of operators to be closed to the Hamiltonian is that they form a Lie algebra. 
In addition the Hamiltonian needs to be a linear combination of operators from the same algebra \cite{Alhassid}.
Finally, the operators have to be closed to the dissipative part ${\cal L}_{h/c}$. 
Most of the solvable models in this review were based on the SU(2) Lie algebra with four operators.
We can chose them to represent the energy $\Op H$,  the identity $\Op I$, and two additional
operators $\Op X$ and $\Op Y$. These operators can associate with coherence since $[\Op H, \Op X] \ne 0,~[\Op H, \Op Y] \ne 0$.
Generically in these models the power is proportional to the coherence ${\cal P} \propto \langle \Op Y \rangle$.
We can speculate that steady state operation of a working engine or refrigerator is minimally characterised by the SU(2)
algebra of three non commuting operators. This is also the case in reciprocating engines \cite{k176,k190,k221}.

The Hamiltonian can be  decomposed to non-commuting local operators which couple to the baths.
The non-local structure influences  the derivation of the master equation. 
A local approach based on equilibrating $\Op H_s$
leads to equations of motion that can violate the second-law of thermodynamics \cite{k122,k275,k278,k281}.
A thermodynamically consistent derivations requires a pre-diagonalization of the system Hamiltonian.
This Hamiltonian defines the system-bath weak coupling approximation leading to the GKS-L completely positive
generator. 

Quantum mechanics introduces dynamics into thermodynamics. 
As a result the laws of thermodynamics have to be reformulated, 
replacing heat to heat currents and work to power. 
The first law at steady state requires that the sum of these currents is zero. 
For the devices considered,  the entropy production is exclusively generated in the baths 
and the sum of all contributions is positive.
All quantum devices should be consistent with the first and second laws of thermodynamics.
Apparent violations point to erroneous derivations of the dynamical equations of motion.

The tricycle template is a universal description for continuous quantum heat engines and refrigerators.
This model incorporates basic thermodynamic ideas within a  quantum formalism. 
The tricycle combines three energy currents from three sources by a nonlinear interaction.
A nonlinear construction is the minimum requirement for all heat engines or refrigerators.
This universality means that the same structure can describe a wide range of quantum devices.
The performance characteristics are given by the choice of working medium and reservoirs.
In the derivation, the reservoirs are characterised by their temperature and correlation functions.
The working medium filters out the channels of heat transfer and power. 

The tradeoff between power and efficiency is a universal characteristic of the quantum devices.
The output power or the cooling current exhibit a turnover with any control parameter \cite{k146}.
For example, when the coupling parameter to the external field is increased first the power or
cooling current increase with it, but then beyond a critical parameter, due to splitting of the energy levels
a decrease in performance appears.  Another turnover phenomena is observed when the coupling parameter to the baths
is increased. Initially increasing the coupling will allow more heat to flow to the device. Above a critical value
dissipation will degrade the coherence and with it the performance.
Optimal power and optimal cooling currents are obtained in a balanced operational point,
far from reversible conditions of maximum efficiency.

The quantum version of the third law originates from the existence of a ground state. 
The third law can be thought of as a limit to the distillation of a pure ground state in a subsystem disentangling it from the environment. 
We can generalize and apply the third law to any pure state of the subsystem since such a state
is accessible from the ground state by a unitary transformation. Unitary transformations  are isoentropic
and therefore allowed by the Nernst heat theorem at $T_c=0$. 
The quantum dilemma of cooling  to a pure state stems from the fact that a
system-bath interaction is required to induce a change of entropy of the system. However, such an interaction generates
system-bath entanglement so that the system cannot be in a pure state. Approaching the absolute zero temperature
is a delicate manoeuvre. The system-bath coupling has to be reduced while the cooling takes place, 
eventually vanishing when $T_c \to 0$. As a result the means to cool are exhausted before the target is reached \cite{k94,segal13}.
The universal behaviour of all quantum refrigerators as $T_c \to 0$ emphasises this point.

How do quantum phenomena such as coherence and entanglement influence performance? 
This issue can be separated
into  global effects which include the reservoirs and internal quantum effects within the device. The description
of the heat transport into the device by the  GKS-L Master equation is the thermodynamic equivalent of an isothermal partition.
In this description the system and bath are not entangled \cite{k281}. 
The issue of internal quantum effects is more subtle.
Internal coherence is necessary for operation in quantum driven devices. 
For quantum autonomous absorption refrigerators it has been recently claimed that, although present, coherence is not determinant in the engine operation \cite{palao13b}. 
Other papers claim the opposite \cite{scully03,popescu13,givonati2013}. Entanglement has been related to extraction of additional work 
\cite{alicki13,karen2013,popescu13b}. These issues require further study.

For completeness we would like to mention what was excluded from the present review: Reciprocating heat engines and refrigerators.
Significant insight can be gained from the analysis of such devices. The template is the quantum Otto four stroke cycle where
the unitary adiabatic branches are separated from the heat transport 
\cite{k85,k87,k116,k221,mahler08,jahnkemahler08,allahmahler08,heJ09,allahaverdian2013,wang13b,wang13}. 
The analysis of reciprocating engine is simplified due to this separation. 
The disadvantage is that the timing of the branches is determined externally.
The review of reciprocating engines and refrigerators will be treated in a separate publication.

Finally, we can conclude from this review that quantum devices when analysed according to the laws of quantum mecahnics
are consistent with a dynamical interpretation of the laws of thermodynamics.

\paragraph{Acknowledgments: }
This review was supported by the Israel Science Foundation. 
We want to thank Tova Feldmann, Eitan Geva, Jose P. Palao, Jeff Gordon, Lajos Diosi, Peter Salamon, 
Gershon Kuritzky  and Robert Alicki for sharing their wisdom.

\break


\begin{thebibliography}{100}
\expandafter\ifx\csname natexlab\endcsname\relax\def\natexlab#1{#1}\fi

\bibitem{carnot}
Carnot S. 1824.
\newblock \textit{{R\'eflections sur la Puissance Motrice du Feu et sur les
  Machines propres \`{a} D\'{e}velopper cette Puissance}}.
\newblock Paris: Bachelier

\bibitem{curzon75}
Curzon F, Ahlborn B. 1975.
\newblock Efficiency of a carnot engine at maximum power output.
\newblock \textit{Am. J. Phys.} 43:22

\bibitem{salamon01}
Salamon P, Nulton J, Siragusa G, Andersen TR, Limon A. 2001.
\newblock Principles of control thermodynamics.
\newblock \textit{Energy} 26:307--319

\bibitem{landsberg56}
Landsberg PT. 1956.
\newblock Foundations of thermodynamics.
\newblock \textit{Rev. Mod. Phys.} 28:363

\bibitem{einstein05}
Einstein A. 1905.
\newblock "\"uber einen die erzeugung und verwandlung des lichtes betreffenden
  heuristischen gesichtspunkt (on a heuristic viewpoint concerning the
  production and transformation of light)".
\newblock \textit{Annalen der Physik} 17:132

\bibitem{scovil59}
Geusic J, Schulz-DuBios E, Scovil H. 1959.
\newblock Three-level masers as heat engines.
\newblock \textit{Phys. Rev. Lett.} 2:262

\bibitem{geusic67}
Geusic J, du~Bois EOS, Grasse RD, Scovil HE. 1967.
\newblock Quantum equivalence of the carnot cycle.
\newblock \textit{Phys. Rev.} 156:343

\bibitem{k24}
Kosloff R. 1984.
\newblock {A Quantum Mechanical Open System as a Model of a Heat Engine.}
\newblock \textit{J. Chem. Phys.} 80:1625--1631

\bibitem{k122}
Geva E, Kosloff R. 1996.
\newblock {The Quantum Heat Engine and Heat Pump: An Irreversible Thermodynamic
  Analysis of The Three-Level Amplifier}.
\newblock \textit{J. Chem. Phys.} 104:7681--7698

\bibitem{partovi89}
Partovi M. {1989}.
\newblock {Quantum thermodynamics}.
\newblock \textit{Physics Letters A} 137:440

\bibitem{k156}
Kosloff R, Geva E, Gordon JM. 2000.
\newblock {The quantum refrigerator in quest of the absolute zero}.
\newblock \textit{J. Appl. Phys.} 87:8093--8097

\bibitem{k169}
Palao JP, Kosloff R, Gordon JM. 2001.
\newblock {Quantum thermodynamic cooling cycle}.
\newblock \textit{Phys. Rev. E} 64:056130--8

\bibitem{lloyd}
Lloyd S. 1997.
\newblock Quantum-mechanical maxwellÕs demon.
\newblock \textit{Phys. Rev. A} 56:3374

\bibitem{he02}
He J, Chen J, Hua B. 2002.
\newblock Quantum refrigeration cycles using spin-$\frac {1}{2}$ systems as
  working substance.
\newblock \textit{Phys. Rev. E} 65:036145

\bibitem{lieb99}
Lieb EH, Yngvason J. 1999.
\newblock The physics and mathematics of the second law of thermodynamics.
\newblock \textit{Phys. Rev.} 310:1

\bibitem{bender}
Bender CM, Brody DC, Meister BK. 2002.
\newblock Entropy and temperature of a quantum carnot engine.
\newblock \textit{Proceedings of the Royal Society of London. Series A:
  Mathematical, Physical and Engineering Sciences} 458:1519

\bibitem{kieu04}
Kieu TD. 2004.
\newblock The second law, maxwell's demon, and work derivable from quantum heat
  engines.
\newblock \textit{Phys. Rev. Lett.} 93:140403

\bibitem{segal06}
Segal D, Nitzan A. 2006.
\newblock Molecular heat pump.
\newblock \textit{Phys. Rev. E} 73:026109

\bibitem{bushev06}
Bushev P, Rotter D, Wilson A, Dubin F, Becher C, et~al. 2006.
\newblock Feedback cooling of a single trapped ion.
\newblock \textit{Phys. Rev. Lett.} 96:60010

\bibitem{nori07}
Quan HT, Liu YX, Sun CP, Nori F. {2007}.
\newblock { Quantum thermodynamic cycles and quantum heat engines }.
\newblock \textit{Phys. Rev. E} 76:031105

\bibitem{erez08}
Boukobza E, Tannor DJ. 2008.
\newblock Thermodynamic analysis of quantum light purification.
\newblock \textit{Phys. Rev. A} 78:013825

\bibitem{mahler08}
Birjukov J, Jahnke T, Mahler G. 2008.
\newblock Quantum thermodynamic processes: a control theory for machine cycles.
\newblock \textit{Eur. Phys. J. B} 64:105

\bibitem{jahnkemahler08}
Jahnke T, Birjukov J, Mahler G. 2008.
\newblock On the nature of thermodynamic extremum principles: The case of
  maximum efficiency and maximum work.
\newblock \textit{Ann.Phys.} 17:88

\bibitem{allahmahler08}
Allahverdyan AE, Johal RS, Mahler G. 2008.
\newblock Work extremum principle: Structure and function of quantum heat
  engines.
\newblock \textit{Phys. Rev. E} 77:041118

\bibitem{segal09}
Segal D. 2009.
\newblock Vibrational relaxation in the kubo oscillator: Stochastic pumping of
  heat.
\newblock \textit{J. Chem. Phys.} 130:134510

\bibitem{he09}
Wang H, Liu S, He J. 2009.
\newblock Thermal entanglement in two-atom cavity QED and the entangled quantum
  otto engine.
\newblock \textit{Phys. Rev. E} 79:041113

\bibitem{nori09}
Maruyama~K, Nori~F, Verdal V. {2009}.
\newblock { Colloquium: The physics of Maxwell's demon and information }.
\newblock \textit{Rev. Mod. Phys.} 81:1

\bibitem{casati13}
Beneti G, Casati G, Prosen T, Saito K. {2013}.
\newblock { Colloquium: Funduamental aspects of steady state heat to work conversion }.
\newblock \textit{Rev. Mod. Phys.} to appear.

\bibitem{mahlerbook}
Gemmer J, Michel M, Mahler G. 2009.
\newblock \textit{{Quantum Thermodynamics}}.
\newblock Springer

\bibitem{olshani12}
{Olshanii} M. 2012.
\newblock {Geometry of quantum observables and thermodynamics of small
  systems}.
\newblock \textit{ArXiv e-prints:1208.0582}

\bibitem{janet13}
Anders J, Giovannetti V. {2013}.
\newblock {Thermodynamics of discrete quantum processes}.
\newblock \textit{New Jour. of Phys.} 15:033022

\bibitem{k281}
Kosloff R. 2013.
\newblock {Quantum Thermodynamics: A Dynamical Viewpoint}.
\newblock \textit{Entropy} 15

\bibitem{jarzynski13}
Mandal D, Quan HT, Jarzynski C. {2013}.
\newblock { MaxwellÕs Refrigerator: An Exactly Solvable Model }.
\newblock \textit{Phys. Rev. Lett.} 111:030602

\bibitem{allahaverdian2013}
Allahverdyan AE, Hovhannisyan KV, Melkikh AV, Gevorkian SG. 2013.
\newblock Carnot cycle at finite power: Attainability of maximal efficiency.
\newblock \textit{Phys. Rev. Lett.} 111:050601

\bibitem{erez13}
Boukobza E, Ritsch H. 2013.
\newblock Breaking the carnot limit without violating the second law: A
  thermodynamic analysis of off-resonant quantum light generation.
\newblock \textit{Phys. Rev. A} 87:063845

\bibitem{k275}
Levy A, Alicki R, Kosloff R. 2012.
\newblock {Quantum refrigerators and the third law of thermodynamics}.
\newblock \textit{Phys. Rev. E} 85:061126

\bibitem{popescu10}
Linden N.~Popescu S. Skrzypczyk P. 2010.
\newblock "how small can thermal machines be? towards the smallest possible
  refrigerato".
\newblock \textit{Phys. Rev. Lett.} 105:130401

\bibitem{palao13}
Luis A.~Correa Jose P.~Palao GA, Alonso D. 2013.
\newblock { Performance bound for quantum absorption refrigerators }.
\newblock \textit{Phys. Rev. E} 87:042131

\bibitem{paz}
Martinez EA, Paz JP. 2013.
\newblock Dynamics and thermodynamics of linear quantum open systems.
\newblock \textit{Phys. Rev. Lett.} 110:130406

\bibitem{clausius1850}
Calusius R. {1850}.
\newblock { "Ueber Die Bewegende Kraft Der Wrme Und Die Gesetze, Welche Sich
  Daraus Fr Die Wrmelehre Selbst Ableiten Lassen"}.
\newblock \textit{Annalen der Physik} 79:68

\bibitem{lindblad76}
Lindblad G. 1976.
\newblock On the generators of quantum dynamical semigroups.
\newblock \textit{Comm. Math. Phys.} 48:119

\bibitem{kossakowski76}
Gorini V, Kossakowski A, Sudarshan E. 1976.
\newblock Completely positive dynamical semigroups of n-level systems.
\newblock \textit{J. Math. Phys.} 17:821

\bibitem{sphon}
Spohn H, Lebowitz J. 1979.
\newblock Irreversible thermodynamics for quantum systems weakly coupled to
  thermal reservoirs.
\newblock \textit{Adv. Chem. Phys.} 38:109

\bibitem{k23}
Kosloff R, Ratner M. 1984.
\newblock {Beyond Linear Response: Lineshapes for Coupled Spins or Oscillators
  via Direct Calculation of Dissipated Power.}
\newblock \textit{J. Chem. Phys.} 80:2352--2362

\bibitem{davis74}
Davis EB. {1974}.
\newblock {Markovian Master Equations}.
\newblock \textit{Comm. Math. Phys.} {39}:{91--110}

\bibitem{k114}
Geva E, Kosloff R, Skinner J. 1995.
\newblock {On the relaxation of a two-level system driven by a strong
  electromagnetic field}.
\newblock \textit{J. Chem. Phys.} 102:8541--8561

\bibitem{alicki2013}
Szczygielski K, Gelbwaser-Klimovsky D, Alicki R. {2013}.
\newblock {Markovian master equation and thermodynamics of a two-level system
  in a strong laser field}.
\newblock \textit{Phys. Rev. E} 87:012120

\bibitem{gorini76}
Gorini V, Kossakowski A. 1976.
\newblock N-level system in contact with a singular reservoir.
\newblock \textit{J. Math. Phys.} 17:1298

\bibitem{cohen87}
{C. Cohen-Tannoudji, J. A. Dupont-Roc, G. Grynberg}. {1987}.
\newblock \textit{{ "Atom-Photon Interaction"}}.
\newblock Wiely

\bibitem{lamb64}
Lamb W. {1964}.
\newblock {Theory of an Optical Maser}.
\newblock \textit{Phys. Rev.} 134:1429

\bibitem{louisell}
Louisell WH. 1990.
\newblock \textit{{Quantum Statistical Properties of Radiation}}.
\newblock Wiley

\bibitem{digonnet90}
Digonnet MJF. {1990}.
\newblock { Closed-Form Expressions for the Gain in Three- and Four-Level Laser
  Fibers }.
\newblock \textit{IEEE J. Quantum Electrinics} 26:1788

\bibitem{scully03}
Scully MO, Zubairy MS, Agarwal GS, Walther H. 2003.
\newblock Extracting work from a single heat bath via vanishing quantum
  coherence.
\newblock \textit{Science} 299:862

\bibitem{scully10}
Scully MO. 2010.
\newblock Quantum photocell: Using quantum coherence to reduce radiative
  recombination and increase efficiency.
\newblock \textit{Phys. Rev. Lett.} 104:207701

\bibitem{scully11}
Svidzinsky AA, Dorfman KE, Scully MO. 2011.
\newblock Enhancing photovoltaic power by fano-induced coherence.
\newblock \textit{Phys. Rev. A} 84:053818

\bibitem{scully11b}
Scully MO, Chapin KR, Dorfman KE, Kim MB, Svidzinsky A. 2011.
\newblock Quantum heat engine power can be increased by noise-induced
  coherence.
\newblock \textit{Proc. Natl. Acad. Sci. U S A} 108:15097

\bibitem{rahav12}
Harbola U, Rahav S, Mukamel S. {2012}.
\newblock {Quantum heat engines: A thermodynamic analysis of power and
  efficiency}.
\newblock \textit{Eur. Phys. Lett.} {99}:{50005}

\bibitem{mukamel12}
Rahav S, Harbola U, Mukamel S. {2012}.
\newblock {Heat fluctuations and coherences in a quantum heat engines}.
\newblock \textit{Phys. Rev. A} {86}:{043843}

\bibitem{harbola13}
Goswami HP, Harbola U. {2013}.
\newblock {Thermodynamics of quantum heat engines }.
\newblock \textit{Phys. Rev. A} 88:013842

\bibitem{gershon13}
D.~Gelbwaser-Klimovsky RA, Kurizki G. 2013.
\newblock Minimal universal quantum heat machine.
\newblock \textit{Phys. Rev. E} 87:012140

\bibitem{erez07}
Boukobza E, Tannor DJ. 2007.
\newblock Three-level systems as amplifiers and attenuators: A thermodynamic
  analysis.
\newblock \textit{Phys. Rev. Lett.} 98:240601

\bibitem{alicki13}
Alicki R, Fannes M. {2013}.
\newblock { Entanglement boost for extractable work from ensembles of quantum
  batteries }.
\newblock \textit{Phys. Rev. E} 87:042123

\bibitem{gelbwaser2013}
{Gelbwaser-Klimovsky} D, {Alicki} R, {Kurizki} G. 2013.
\newblock {How much work can a quantum device extract from a heat engine?}
\newblock \textit{ArXiv e-prints: 1302.3468}

\bibitem{nerst06}
Nernst W. 1906.
\newblock {Ueber die Berechnung chemischer Gleichgewichte aus thermischen
  Messungen}.
\newblock \textit{{Nachr. Kgl. Ges. Wiss. G\"ott.}} 1:40

\bibitem{nerst06b}
Nernst W. 1906.
\newblock {U\"ber die Beziehung zwischen Wa\"rmeentwicklung und maximaler
  Arbeit bei kondensierten Systemen}.
\newblock \textit{{er. Kgl. Pr. Akad. Wiss.}} 52:933

\bibitem{nerst18}
Nernst W. 1918.
\newblock \textit{{The theoretical and experimental bases of the New Heat
  Theorem Ger., Die theoretischen und experimentellen Grundlagen des neuen
  Wa\"rmesatzes}}.
\newblock Halle: W. Knapp

\bibitem{Fowler39}
Fowler RH, Guggenheim EA. 1939.
\newblock \textit{{Statistical Thermodynamics}}.
\newblock Cambridge university press

\bibitem{landsberg89}
Landsberg PT. 1989.
\newblock A comment on nernst's theorem.
\newblock \textit{J. Phys A: Math.Gen.} 22:139

\bibitem{belgiorno03}
Belgiorno F. 2003.
\newblock Notes on the third law of thermodynamics i.
\newblock \textit{J. Phys A: Math.Gen.} 36:8165

\bibitem{belgiorno03b}
Belgiorno F. 2003.
\newblock Notes on the third law of thermodynamics ii.
\newblock \textit{J. Phys A: Math.Gen.} 36:8195

\bibitem{k278}
{Levy A, Alicki R, Kosloff R}. 2012.
\newblock {Comment on Cooling by Heating: Refrigeration Powered by Photons"}.
\newblock \textit{Phys. Rev. Lett.} 109:248901

\bibitem{geustic59}
Geusic J, Bois E, De~Grasse R, Scovil H. 1959.
\newblock Three level spin refrigeration and maser action at 1500 mcsec.
\newblock \textit{J. App. Phys.} 30:1113

\bibitem{pikuji63}
Tsujikawa I, Murao T. {1962}.
\newblock {Possibility of Optical Cooling of Ruby}.
\newblock \textit{J. Phys. Soc. Jpn.} 18:503

\bibitem{Dousmanis64}
G.~C.~Dousmanis C. W.~Mueller HN, Petzinger KG. 1964.
\newblock Evidence of refrigerating action by means of photon emission in
  semiconductor diodes.
\newblock \textit{Phys. Rev.} 133:A316

\bibitem{geustic68}
Kushida T, Geusic JE. 1968.
\newblock Optical refrigeration in nd-doped yittrium aluminum garnet.
\newblock \textit{Phys. Rev. Lett.} 20:1172

\bibitem{wineland75}
Wineland DJ, Dehmelt H. {1975}.
\newblock {}.
\newblock \textit{Bull. Am. Phys. Soc.} 20:637

\bibitem{hansch75}
H\"ansch TW, Schawlow AL. {1975}.
\newblock {Cooling of Gases by Laser Radiation}.
\newblock \textit{Opt. Commun.} 13:368

\bibitem{philips88}
Lett PD, Watts RN, Westbrook CI, Phillips WD, Gould PL, Metcalf HJ. {1988}.
\newblock Observation of atoms laser cooled below the doppler limit.
\newblock \textit{Phys. Rev. Lett.} 61:169

\bibitem{chu89}
Shevy Y, Weiss DS, Ungar PJ, Chu S. {1989}.
\newblock { Bimodal speed distributions in laser-cooled atoms }.
\newblock \textit{Phys. Rev. Lett.} 62:1118

\bibitem{cohen89}
Dalibard J, Cohen-Tannoudji C. {1989}.
\newblock { "Laser cooling below the Doppler limit by polarization gradients:
  simple theoretical models"}.
\newblock \textit{J. Opt. Soc. Am. B} 79:2023

\bibitem{steck2003}
Steck DA. {2003}.
\newblock { Sodium D Line Data}\\ 
\newblock \textit{steck.us/alkalidata/sodiumnumbers.1.6.pdf} 

\bibitem{jeff00}
Gordon JM, Ng KC. 2000.
\newblock \textit{Cool Thermodynamics}.
\newblock Cambridge International Science Publishing

\bibitem{szilard1926}
Einstein A, Szil\'ard L. 1930.
\newblock "refrigeration".
\newblock \textit{"US patent No 1,781,541"}

\bibitem{mari2012}
Mari A, Eisert J. 2012.
\newblock Cooling by heating: Very hot thermal light can significantly cool
  quantum systems.
\newblock \textit{Phys. Rev. Lett.} 108:120602

\bibitem{vandebrock12}
Cleuren B, Rutten B, Van~den Broeck C. {2012}.
\newblock {Cooling by Heating: Refrigeration Powered by Photons}.
\newblock \textit{Phys. Rev. Lett.} {108}:{120603}

\bibitem{givonati2013}
Venturelli D, Fazio R, Giovannetti V. 2013.
\newblock Minimal self-contained quantum refrigeration machine based on four
  quantum dots.
\newblock \textit{Phys. Rev. Lett.} 110:256801

\bibitem{pekola12}
Pekola JP, Hekking FWJ. {2007}.
\newblock {Normal-Metal-Superconductor Tunnel Junction as a Brownian
  Refrigerator}.
\newblock \textit{Phys. Rev. Lett.} {98}:{210604}

\bibitem{palao13b}
{Correa} LA, {Palao} JP, {Alonso} D, {Adesso} G. 2013.
\newblock {Quantum-enhanced absorption refrigerators}.
\newblock \textit{ArXiv e-prints:1308.4174}

\bibitem{k272}
{Levy A, Kosloff R}. 2012.
\newblock {Quantum absorption refrigerator}.
\newblock \textit{Phys. Rev. Lett.} 108:070604

\bibitem{popescu11}
Skrzypczyk P~Brunner N, Popescu S. 2011.
\newblock "the smallest refrigerators can reach maximal efficiency".
\newblock \textit{J. Phys A: Math.Gen.} 44:492002

\bibitem{popescu12}
Brunner N, Linden N, Popescu S, Skrzypczyk P. 2012.
\newblock "virtual qubits, virtual temperatures, and the foundations of
  thermodynamics".
\newblock \textit{Phys. Rev. E} 85:051117

\bibitem{breuer}
Breuer HP, Petruccione F. 2002.
\newblock \textit{Open quantum systems}.
\newblock Oxford university press

\bibitem{luczka91}
\'Luczka J, Niemeic M. 1991.
\newblock "a master equation for quantum systems driven by poisson white
  noise".
\newblock \textit{J. Phys A: Math.Gen.} 24:L1021

\bibitem{alicki06}
Alicki R, Lidar DA, Zanardi P. 2006.
\newblock Internal consistency of fault-tolerant quantum error correction in light
 of rigorous derivations of the quantum Markovian limit .
\newblock \textit{Phys. Rev. A} 73:052311

\bibitem{gershon12}
Kol{\'a}{\v{r}} M, Gelbwaser-Klimovsky D, Alicki R, Kurizki G. 2012.
\newblock Quantum bath refrigeration towards absolute zero: unattainability
  principle challenged.
\newblock \textit{Phys. Rev. Lett.} 108:090601

\bibitem{Alhassid}
Alhassid Y, Levine R. 1978.
\newblock Exact conditions for the preservation of a canonical distribution in
  markovian relaxation processes.
\newblock \textit{Phys. Rev. A} 18:89

\bibitem{k176}
Kosloff R, Feldmann T. 2002.
\newblock {A Discrete Four Stroke Quantum Heat Engine Exploring the Origin of
  Friction.}
\newblock \textit{Phys. Rev. E} 65:055102

\bibitem{k190}
Feldmann T, Kosloff R. 2003.
\newblock {The Quantum Four Stroke Heat Engine: Thermodynamic Observables in a
  Model with Intrinsic Friction}.
\newblock \textit{Phys. Rev. E} 68:016101

\bibitem{k221}
Rezek Y, Kosloff R. 2006.
\newblock {Irreversible performance of a quantum harmonic heat engine}.
\newblock \textit{New J. Phys.} 8:83

\bibitem{k146}
{Ashkenazi G, Kosloff R and Ratner MA}. 1999.
\newblock {Photoexcited electron transfer: Short-time dynamics and turnover
  control by dephasing, relaxation, and mixing}.
\newblock \textit{J. Am. Chem. Soc.} 121:3386--3395

\bibitem{k94}
Bartana A, Kosloff R, Tannor DJ. 1993.
\newblock {Laser Cooling of Molecular Internal Degrees of Freedom by a Series
  of Shaped Pulses}.
\newblock \textit{J. Chem. Phys.} 99:196--210

\bibitem{segal13}
{Wu L., Segal D. and Brumer P.} {2013}.
\newblock { No-go theorem for ground state cooling given initial system-thermal
  bath factorization }.
\newblock \textit{Sci Rep.} 3:1824

\bibitem{popescu13}
{Skrzypczyk} P, {Short} AJ, {Popescu} S. 2013.
\newblock {Extracting work from quantum systems}.
\newblock \textit{ArXiv e-prints: 1302.2811}

\bibitem{karen2013}
{Hovhannisyan} KV, {Perarnau-Llobet} M, {Huber} M, {Ac{\'{\i}}n} A. 2013.
\newblock {The role of entanglement in work extraction}.
\newblock \textit{ArXiv e-prints:1303.4686}

\bibitem{popescu13b}
{Brunner} N, {Huber} M, {Linden} N, {Popescu} S, {Silva} R, {Skrzypczyk} P.
  2013.
\newblock {Entanglement enhances performance in microscopic quantum fridges}.
\newblock \textit{ArXiv e-prints: 1305.6009}

\bibitem{k85}
Geva E, Kosloff R. 1992.
\newblock {A Quantum Mechanical Heat Engine Operating in Finite Time. A Model
  Consisting of Spin $~half~$ Systems as The Working Fluid}.
\newblock \textit{J. Chem. Phys.} 96:3054--3067

\bibitem{k87}
Geva E, Kosloff R. 1992.
\newblock {On the Classical Limit of Quantum Thermodynamics in Finite Time}.
\newblock \textit{J. Chem. Phys.} 97:4398--4412

\bibitem{k116}
Feldmann T, Geva E, Kosloff R, Salamon P. 1996.
\newblock {Heat Engines in Finite Time Governed by Master Equations}.
\newblock \textit{{Am. J. Phys.}} 64:485--492

\bibitem{heJ09}
He~JiZhou HX, Wei T. 2009.
\newblock "the performance characteristics of an irreversible quantum otto
  harmonic refrigeration cycle".
\newblock \textit{Science in China Series G-Phys. Mech. \& Ast.} 52:{1317}

\bibitem{wang13b}
Wang H. {2013}.
\newblock Quantum-mechanical brayton engine working with a particle in a
  one-dimensional harmonic trap.
\newblock \textit{Physica Scripta} 87:055009

\bibitem{wang13}
Wang R, Wang J, He J, Ma Y. 2013.
\newblock { Efficiency at maximum power of a heat engine working with a
  two-level atomic system }.
\newblock \textit{Phys. Rev. E} 87:042119

\end{thebibliography}
\end{document}